\documentclass[10pt,preprint]{aastex}

\newcommand{\myemail}{ipillitteri@cfa.harvard.edu}
\newcommand{\fx}{$F_\mathrm X$}
\newcommand{\soxs}{{\em SOXS} }
\newcommand{\lx}{$L_\mathrm X$}
\newcommand{\lxu}{erg s$^{-1}$}
\newcommand{\fxu}{erg s$^{-1}$ cm$^{-2}$}

\newcommand{\xmm}{{\em XMM-Newton}}
\newcommand{\spitzer}{{\em Spitzer}}
\newcommand{\chandra}{{\em Chandra}}
\newcommand{\srcnr}{{ $1060$\ }}
\newcommand{\iori}{{$\iota$~Ori}}

\shorttitle{X-ray sources in the \soxs survey.}
\shortauthors{Pillitteri et al.}

\begin{document}

\title{An X-rays Survey of the Young Stellar Population of the Lynds 1641 and Iota Orionis Regions.}

\author{I. Pillitteri\altaffilmark{1} 
\and S. J. Wolk\altaffilmark{1}
\and  S. T. Megeath\altaffilmark{2} 
\and L. Allen\altaffilmark{3} 
\and J. Bally\altaffilmark{4}
\and Marc Gagn\'e\altaffilmark{5}  
\and R. A. Gutermuth\altaffilmark{6} 
\and L. Hartman\altaffilmark{7}
\and G. Micela\altaffilmark{8}
\and P. Myers\altaffilmark{1}
\and J. M. Oliveira\altaffilmark{9}
\and S. Sciortino\altaffilmark{8}
\and F. Walter\altaffilmark{1}
\and L. Rebull\altaffilmark{10}
\and J. Stauffer\altaffilmark{10}
}
\affil{SAO--Harvard Center for Astrophysics, 60 Garden St, Cambridge MA 02138 -- USA} %
\and
\affil{ Department of Physics \& Astronomy, University of Toledo, OH -- USA} %
\and
\affil{National Optical Astronomy Observatory -- USA} %
\and
\affil{University of Colorado, Boulder, CO -- USA } %
\and
\affil{Department of Geology \& Astronomy, West Chester University, West Chester, PA -- USA} %
\and
\affil{Dept. of Astronomy, University of Massachusetts, Amherst, MA 01003  -- USA} %
\and
\affil{University of Michigan, Ann Arbor, MI -- USA} %
\and
\affil{INAF - Osservatorio Astronomico di Palermo -- Italy} %
\and
\affil{School of Physical \& Geographical Sciences, Lennard-Jones
Laboratories, Keele University, Staffordshire ST5 5BG -- UK  } %
\and
\affil{CALTECH, Pasadena, CA, 91125 -- USA } %

\email{\myemail}

\begin{abstract}
{ We present an \xmm\ survey of the part of Orion A cloud south of the Orion Nebula.  
This survey includes the { Lynds 1641 (L1641) dark cloud}, a region of the Orion~A cloud with very 
few massive stars and hence a relatively  low ambient UV flux, and the region around 
the O9 III star $\iota$ Orionis.  In addition to proprietary data, 
we used archival XMM data of the Orion Nebula Cluster (ONC) to extend our analysis to a major 
fraction of the Orion~A cloud.

We have detected 1060 X-ray sources in L1641 and  \iori\ region. About
94\% of the sources have 2MASS \& {\em Spitzer} counterparts, 204 and 23 being Class II and 
Class I or protostars objects, respectively. 
In addition, we have identified 489 X-ray sources as counterparts to Class III candidates, 
given they are bright in X-rays and appear as normal photospheres at mid-IR wavelengths. 
The remaining 205 X-ray sources are likely distant AGNs or other galactic 
sources not related to Orion~A.

We find that Class III candidates appear more concentrated in two main clusters in L1641. 
The first cluster of Class III stars is found toward the northern part of L1641, concentrated
around \iori. The stars in this cluster are more evolved than those in the Orion Nebula. 
We estimate a distance of  $300-320$~pc for this cluster and  thus it is closer than the Orion~A cloud. 

Another cluster rich in Class III stars is located in L1641 South and appears to be a slightly older 
cluster embedded in the Orion A cloud.  Furthermore, other evolved Class III stars are found north
of the ONC toward NGC 1977.}
\end{abstract}

\section{Introduction}
Stars form by the fragmentation and collapse of dense gas within molecular clouds 
\citep{Carpenter00,Odell2001,Allen07,Bally08}. Surveys of the population of young stars within a cloud
have a crucial role in determining the history and the efficiency of star formation. 
A multi-wavelength approach is invaluable for this purpose, because of the intense X-ray and IR radiation 
typical of young stars. 

The first phases of stellar formation are characterized by strong 
infrared (IR) and X-ray band emission. The IR excess comes from the circumstellar
disk and the inner envelope that surround the protostellar core ({\em Class 0--I} and {\em flat spectrum} 
objects). The disk is present in the subsequent stages ({\em Class II} objects, 
or {\em Classical T-Tauri} stars; cTTs) until accretion ends and gas-rich disks are dissipated by 
stellar wind  ({\em Class III} objects or {\em Weak T-Tauri} stars, wTTs) or eventually condense into 
planetesimals. 
X-rays are thought to come from a scaled up version of the solar corona, which should form quite 
early in the pre-main sequence (PMS) phase \citep{Favatarev03,Feigelson1999}. 
In addition, a soft X-ray component can arise from shocks formed by
accretion process or the interaction of outflows and the circumstellar material 
\citep{Favata05,Guedel08, Brickhouse2010, Argiroffi2011}. 
X-rays are thus a powerful tracer of youth, allowing us to identify young stars after their IR signatures 
have disappeared.
Multi-wavelength surveys can map the distribution of stars as a function of the evolutionary class, thus enabling
studies of the star formation history within molecular clouds.

One of the best studied star forming regions is the Orion Molecular Cloud (OMC) complex,
located in the Orion OB1 association.
Comprehensive overviews of the morphology of the OMC and the mechanisms
that have shaped its complex star formation history are given by \citet{Bally08}.
Briefly, within the OMC two main giant clouds are identified, Orion A and Orion B,
extended in declination in the range roughly from $-10\deg$ to $-5\deg$ (Orion A) and from $-4\deg$ to $+4\deg$ 
(Orion B).
Furthermore, several OB associations are present and their subgroups span an age range 
between $<1$ Myr and 10-12 Myr, for a total population estimated between 5000 and 20000 stars.

We have focused our attention on the southern part of the OMC and the Orion A cloud,
and in particular the region south of the { Orion Nebula Cluster (ONC)} that includes
the gas filaments of Lynds 1641 North and South (L1641 N, L1641 S). This region contains also 
$\iota$ Orionis and the cluster NGC~1980 loosely associated with it. 
\iori\ seems to be closer than L1641 and the ONC by $\sim10-50$ pc \citep{Bally08} because of 
the absence of any reflection effect on the background cloud. 
The Orion A molecular cloud is less dense in this region than in the ONC. It has a filamentary 
shape approximately $0.5^\mathrm o \times2.5^\mathrm o$ long (see Fig. \ref{soxsmap}). 
Maps in $^{12}$CO and  $^{13}$CO reveal several dense, small clumps along the filament 
that host small groups of  young stars \citep{Bally87,Strom89,Allen95}.
A gradient of velocity in the gas cloud is detected from CO maps \citep{Bally87} that could
imply a tilt to the cloud and a gradient in its distance decreasing from South to North, 
with the southern part being closer by $\leq10$\%.
In this paper we will assume the distance to the ONC given by \citet[][414 pc]{Menten07}.

X-ray surveys of L1641, obtained with {\em Einstein} \citep{Strom90}, dramatically increased
the number of known young stellar objects compared to previous optical/IR surveys.
More recent observations with \xmm\ and \chandra\ have studied only specific objects, 
like HH1/HH2 objects (with \chandra),  and V883 Ori, V380 Ori (with \xmm). 
Some of { the \xmm\ archival observations} have been taken into account and reanalyzed in this 
paper to complete our survey { (see Table \ref{fields})}.   
With the advent of \spitzer\ and its high sensitivity in the mid-IR, 
the knowledge of this region has been substantially enhanced (\citealp{Megeath2012}, \citealp{Fang09}).

In contrast to the ONC, L1641 does not contain any early B type or O stars \citep{Hsu2013,Allen07}. 
As a consequence, the strong ultraviolet (UV) flux illuminating the circumstellar disks and the 
envelopes of PMS stars of the ONC is almost absent in this region. 
The lack of strong UV and far UV (FUV) fluxes may be important 
for the lifetime of disks, their evaporation, chemistry and, in turn, the evolution 
of the angular momentum of stars, their activity and, finally, the formation of planets
around these stars \citep{Guarcello2010,Wright2012}.

The first step for addressing these issues is to have a complete census of the 
stars at different evolutionary stages of their early evolution. 
For this purpose, we embarked on a {\em Survey of Orion A with XMM-Newton and Spitzer} ({\em SOXS}). 
The goal of \soxs is to detect the relatively bright X-ray sources in the Orion A cloud
and complete the census of more evolved Young Stellar Objects (YSOs) in this region.  
These stars represent a sample of cloud members chosen with minimal {\em a priori} bias 
toward their IR properties.
We can then examine the IR and spatial properties of these sources to ascertain 
the distribution of YSOs within the cloud. 
Fig. \ref{fields} (right panel) shows a map of extinction, overlaid on the \spitzer\ sky coverage
and labels to indicate the different regions that we will discuss in the paper.
The \xmm\ survey covers about 60\% of the sky coverage of the \spitzer\ survey, 
although it comprises the majority of the groups of YSOs detected in IR.
This paper presents the catalog of sources detected in the X-ray survey of \soxs
and their basic X-ray characteristic, as well as clustering metrics of the 
young stellar population in L1641.
The structure of the paper is the following: in \S \ref{observations} we describe 
the data we have acquired and their analysis;
in \S \ref{results} and \ref{ysopop} we report our results, in \S \ref{newclusters} we discuss
the discovery of new groups of Class III stars, and in \S \ref{conclusions} we give
our conclusions.

\section{Observations and data analysis} \label{observations}
\begin{figure*}
\begin{center}
\includegraphics[width=0.62\textwidth]{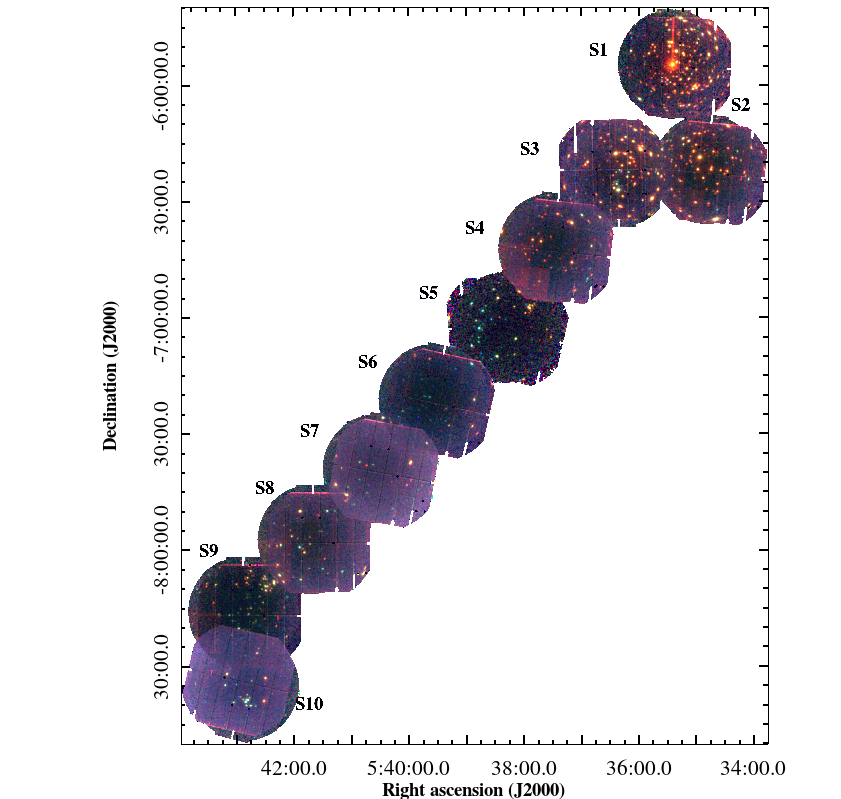}
\includegraphics[width=0.62\textwidth]{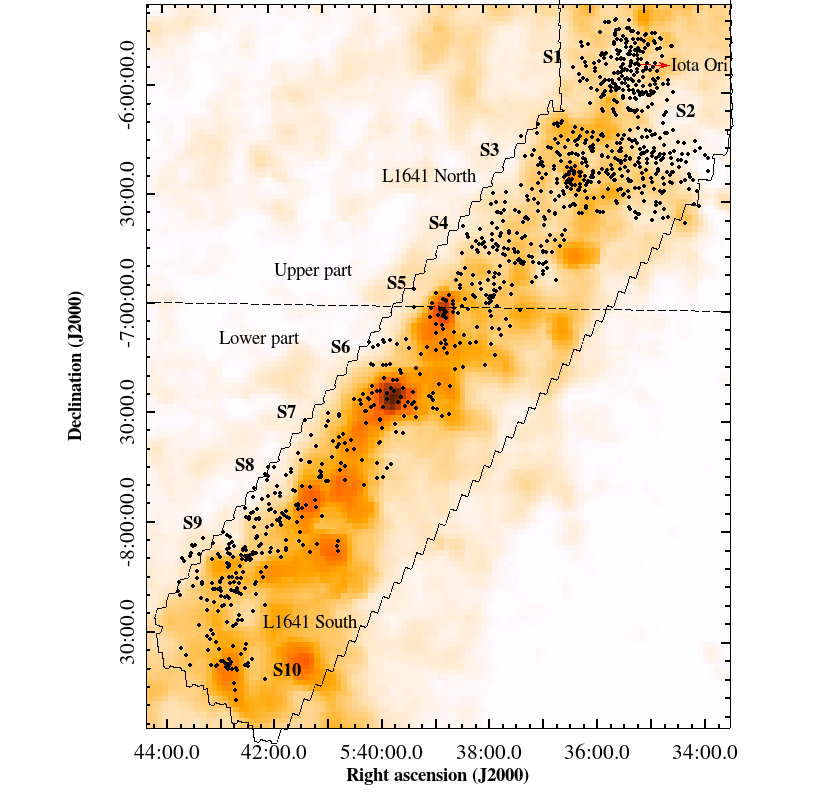}
\end{center}
\caption{\label{soxsmap} Top panel: RGB color image of the mosaic of SOXS fields.
Bands used: Red $= 0.3-1.0$ keV; Green $= 1.0-2.5$; Blue $=2.5-8.0$ keV.
The fields are labeled. 
Bottom panel: %
$A_V$ extinction map from \citet{Gutermuth2011} with the positions of X-ray sources. 
{The scale is linear with darker colors indicating higher $A_V$ values.}
The range of $A_V$ is the same of fig. \ref{nh-av}, approximately from $\sim1$ to $\simeq18$ mag.
The black lines show the extent of the Spitzer field.
About 60\% of the \spitzer\ field of view has been observed by \xmm, however most of the YSOs are found
in the densest part of the cloud.
}
\end{figure*}

The \soxs survey is composed of seven specifically proposed $\sim50$ ks \xmm\ fields, 
to which we added four archival fields in the
same region, south of the ONC. The archival fields are centered on $\iota$ Orionis, V380~Ori and V883~Ori, 
respectively (Fig. \ref{soxsmap}).
As shown in Fig. \ref{fields} (right panel), we have observed in X-rays the dense part of the filament of 
L1641 while \spitzer\ has observed a wider area (\citealp{Megeath2012}, \citealp{Gutermuth2011}).

\begin{figure*}
\begin{center}
\includegraphics[width=0.99\textwidth]{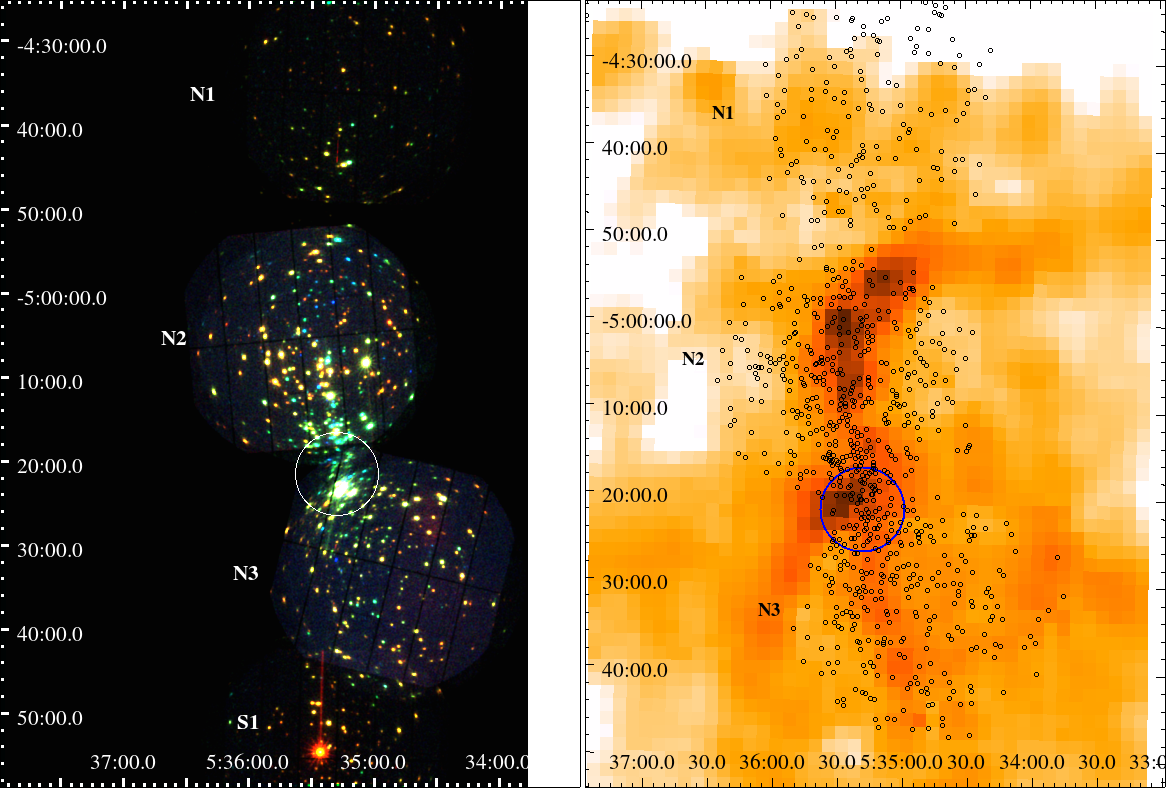}
\end{center}
\caption{\label{northernfields} 
Left: mosaic of EPIC-\xmm\ fields $N1$, $N2$ and $N3$. The circle between fields $N2$ and $N3$ 
indicates the crowded part of the X-ray images, and corresponding approximately to the COUP 
field, that was excluded by our analysis. 
Right panel: 
$A_V$ extinction map adapted from \citet{Gutermuth2011} and positions of detected X-ray sources. 
{The scale is linear with darker colors indicating higher $A_V$ values.}
The range of $A_V$ is approximately from 1 to 18 mag. Colors and symbols are as in fig. \ref{soxsmap}. }
\end{figure*}

Table \ref{fields} gives a log of the observations. The SOXS fields in L1641 are labeled $S1$ to $S10$. 
Typical exposure times are of order $50-60$~ks for the seven \soxs fields and $\iota$~Orionis field,
while exposures are shorter ($\sim15-20$~ks) for the V380~Ori and V883~Ori fields. 
The field of \object{V883 Ori} has been observed twice, with observations taken five months apart.
These two archival observations are called $S5a$ and $S5b$ and are analyzed together to improve the 
count statistics (ObsId \#0205150401 and \#0205150501). The {\em Medium} filter was used to screen
UV contamination for all fields except for \object{$\iota$~Orionis} and 
\object{V380 Ori} fields,  where {\em Thick} and {\em Thin1} filters were used, respectively. 
The differences of effective area among these filters is small above 0.7--0.8 keV where most of 
X-rays from YSOs are detected. We have also considered other archival fields containing the ONC 
(named $N1$, $N2$, $N3$, Fig. \ref{northernfields}), 
to derive X-ray luminosity distributions and for comparison with L1641.

{ The Point Spread Function (PSF) of \xmm\ has a core width of about 5\arcsec. Different from \chandra, 
it remains quite constant at increasing off-axis angles. Its shape becomes less circular and more elongated at 
high off-axis angles. At outer regions of the field of view the sensitivity is mainly affected by the vignetting,
and only at very high off-axis angles by the change of shape of the PSF. In our analysis, this
problem affects the region across fields $N2$ and $N3$ which contains the core of the ONC. This region has
been excluded from the subsequent analysis, whereas in the southern fields $S1$ to $S10$ the source  
confusion is not an issue.}

\begin{table*}
\caption{\label{fields} List of \xmm\ observations. Table contains the field names, observation identifier, 
R.A. and Dec. of the pointings, dates of the observations, { exposure times after GTI filtering}, and 
{\em EPIC} filters used during exposures. For ObsId \#0212481101 only the PN exposure has been analyzed.} 
 \scriptsize
\begin{tabular}{lrcccccl}\hline\hline 
\multicolumn{8}{c}{ Southern fields} \\ 
{Field Name} & {ObsId \#} & { R.A. (J2000)} & {Dec. (J2000)} & {Date} &  {Exp. Time (ks)} & {Filter} & {Notes}\\ \hline 
S1 & 0112660101  &       05:35:25.98     &  -05:54:35.6 &    2001-09-15 &    22.5   &   Thick   & Archive -- $\iota$ Ori\\
S2  & 0503560701  &       05:34:49.90     &  -06:21:53.0 &    2007-09-10 &    55.4   &   Medium  & This program\\
S3 & 0089940301  &       05:36:22.30     &  -06:22:19.0 &    2001-03-18 &    45.8   &   Thin1   & Archive -- V380 Ori\\
S4  & 0503560601  &       05:37:30.20     &  -06:42:03.9 &    2007-09-08 &    55.1   &   Medium  & This program\\
S5a & 0205150401  &       05:38:18.10     &  -07:02:24.9 &    2004-03-27 &     4.6   &   Medium  & Archive -- V883 Ori\\
S5b & 0205150501  &       05:38:18.10     &  -07:02:24.9 &    2004-08-25 &     7.7   &   Medium  & Archive -- V883 Ori\\
S6  & 0503560101  &       05:39:35.10     &  -07:21:09.3 &    2007-08-27 &    48.3   &   Medium  & This program\\
S7  & 0503560401  &       05:40:33.30     &  -07:39:07.7 &    2007-08-31 &    53.0   &   Medium  & This program\\
S8  & 0503560201  &       05:41:41.37     &  -07:57:19.1 &    2007-09-20 &    60.0   &   Medium  & This program\\
S9  & 0503560301  &       05:42:54.00     &  -08:15:52.6 &    2007-09-20 &    41.6   &   Medium  & This program\\
S10 & 0503560501  &       05:42:52.70     &  -08:34:37.7 &    2008-02-23 &    46.5   &   Medium  & This program\\\hline \hline
\multicolumn{8}{c}{ Northern fields} \\ 
{Field Name} & {ObsId \#} & { R.A. (J2000)} & {Dec. (J2000)} & {Date} &  {Exposure Time (ks)} & {Filter} & {Notes}\\ \hline 
N1  & 0049560301  &      05:35:10.81      &  -04:34:06.8 &    2002-09-15 &   26.3    & Thick     & Archive \\
N2a & 0093000101  &      05:35:27.70      &  -05:05:56.5 &    2001-03-25 &   60.0    & Medium    & Archive \\
N2b & 0093000301  &      05:35:27.70      &  -05:05:56.5 &    2001-03-26 &   17.3    & Medium    & Archive \\
N2c & 0134531601  &      05:35:27.70      &  -05:05:56.5 &    2003-03-15 &   20.7    & Medium    & Archive \\
N2d & 0134531701  &      05:35:27.70      &  -05:05:56.5 &    2003-09-15 &   20.9    & Medium    & Archive \\
N3a & 0212480301  &      05:34:44.70      &  -05:33:41.3 &    2005-02-18 &   19.9    & Medium    & Archive -- V1118 Ori\\
N3b & 0212480401  &      05:34:44.70      &  -05:33:41.3 &    2005-03-21 &   19.8    & Medium    & Archive -- V1118 Ori\\
N3c & 0212481101  &      05:34:44.70      &  -05:33:41.3 &    2005-09-08 &   12.4    & Medium    & Archive -- V1118 Ori\\
\hline
\end{tabular}
\end{table*}

\subsection{\xmm\ data analysis.} \label{analysis}
The initial screening and analysis of the data
were performed using {\em SAS} software ver. 8.0. From the {\em Observation Data Files},  
we obtained fits tables of events recorded by the EPIC cameras PN, MOS1, and MOS2 with 
calibrated energies and astrometry. Subsequently, we limited events to those within the $0.3-8.0$ keV band, 
removed the events out of the field of view, and selected only single and double events 
(PATTERN $\leq$ 12). We have also screened the exposure times (good time intervals; GTIs) of 
each field to improve 
the signal-to-noise ratio (SNR) of faint sources in order to effectively detect them.
This step is necessary to remove the intervals of high background due, for example, to solar activity. 
We used the algorithm of \cite{Sciortino01} elaborated by \citet{Damiani2003} 
that maximizes the SNR of the whole image, by assuming that the variability in 
the global light curve is determined by the background 
variability only. We choose the GTIs that realize the maximum SNR and this 
defines a limiting threshold in the count rate. 

We performed source detection with a wavelet-based code derived from an analogous code developed 
for the {\em ROSAT} and {\em Chandra} instruments ({\em Pwxdetect}, see \citealp{Dami97.1,Sciorti2001}).
The version of the code is adapted to detect sources in a list of \xmm\ EPIC images combining the original
unbinned event files in the weighted sum  of MOS and PN images, 
and with the multi-scale approach inherited from its ROSAT predecessor (see also \citealp{Pillitteri2010aa}). 

In order to choose a proper threshold for detection, several cycles simulating source detection 
on background only images have been performed. We examined
the cumulative distribution of the significance of detected spurious sources in these images
and chosen a threshold value to limit false detections to 1-2 per field.
The threshold for significance of source detection was $\ge 4.5\sigma$ of the
local background mean in most cases. In the short exposure fields,the threshold has been chosen slightly lower,
following the results of the simulations.
The observations of the field of 
V883 Ori (ObsId \#0205150401 and \#0205150501) were merged into a single 
observation  prior to analysis, as they are identical pointings.
The final list of X-ray sources has been checked to remove false sources which are 
caused by bright pixels, or spikes around very bright sources. 
Table \ref{detlist} report the first ten out of 1060 sources detected in fields $S1-S10$.
\begin{table}
\caption{\label{detlist}  List of X-ray detected sources. 
We list identifier, coordinates, off-axis distance with respect to the aim point of the observation, 
net total counts (MOS1+MOS2+PN), count rates scaled to MOS sensitivity, sum of exposure times from 
EPIC cameras,  and XMM observation identifier. Counts and count rates are given in the $0.3-8.0$ keV band.
}\small
\begin{tabular}{lccrrrrr}\hline\hline
Id.		 &	R.A. (J2000)	 & 	Dec. (J2000)	 &	Off-axis  & 	Counts		 & 	Rate 		 &	Exp. Time&	ObsId \\
		 &	deg.		 & 	deg.		 &	\arcmin	 &			 & 	ct/ks		 & 	ks	 & 		  \\\hline
 1		 & 	05:33:49.5	 & 	-06:18:05.9	 & 	14.14	 & 	33$\pm$9	 & 	0.74$\pm$0.2	 & 	27.67	 & 	0503560701  \\
 2		 & 	05:33:50.3	 & 	-06:21:31.6	 & 	13.79	 & 	625$\pm$38	 & 	13.69$\pm$0.83	 & 	27.94	 & 	0503560701  \\
 3		 & 	05:33:53.3	 & 	-06:17:11.1	 & 	13.40	 & 	376$\pm$30	 & 	8.15$\pm$0.65	 & 	28.25	 & 	0503560701  \\
 4		 & 	05:33:58.5	 & 	-06:15:14.2	 & 	12.84	 & 	96$\pm$17	 & 	1.73$\pm$0.3	 & 	34.08	 & 	0503560701  \\
 5		 & 	05:34:00.0	 & 	-06:31:14.2	 & 	15.60	 & 	202$\pm$23	 & 	4.51$\pm$0.51	 & 	27.37	 & 	0503560701  \\
 6		 & 	05:34:02.1	 & 	-06:27:07.3	 & 	12.66	 & 	125$\pm$20	 & 	1.49$\pm$0.24	 & 	51.16	 & 	0503560701  \\
 7		 & 	05:34:03.9	 & 	-06:16:03.7	 & 	11.28	 & 	396$\pm$46	 & 	10.14$\pm$1.2	 & 	23.91	 & 	0503560701  \\
 8		 & 	05:34:04.0	 & 	-06:13:30.2	 & 	12.49	 & 	71$\pm$15	 & 	0.86$\pm$0.19	 & 	50.69	 & 	0503560701  \\
 9		 & 	05:34:06.3	 & 	-06:30:46.5	 & 	14.16	 & 	778$\pm$43	 & 	14.12$\pm$0.78	 & 	33.72	 & 	0503560701  \\
 10		 & 	05:34:06.6	 & 	-06:24:13	 & 	10.37	 & 	629$\pm$48	 & 	11.88$\pm$0.9	 & 	32.37	 & 	0503560701  \\\hline
\end{tabular}\\
\footnotesize{First 10 rows are shown, full table is published in electronic format only.}
\end{table}

We defined extraction regions for sources and background 
to select events and produce light curves and spectra 
by means of the SAS task {\em regions}. 
The task defines circular regions for each source and tries to maximize the SNR 
with respect to a local background while avoiding overlap between nearby sources. 
Events, light curves and spectra have been extracted for each source.
Spectra were obtained with the SAS task {\em especget}. For each source, the task creates the
source and background spectra, calculates the ancillary response file (ARF) and
writes the area scaling factors for background and source spectra into the respective 
headers of the files. 
For PN spectra, it associates and writes in the spectrum file the correct response matrix (RMF) 
from the list of pre-calculated response files in the \xmm\ archive, 
while for MOS, due to the peculiar form of the RMF
and the different contributions to the RMF, ad hoc RMFs have been calculated 
for each source by using the task {\em rmfgen}.
For sources with more than 500 total counts (232 out of 1060) as determined from the wavelet analysis, 
spectra from MOS1, MOS2 and PN were grouped to have at least 25 counts per bin and fit simultaneously 
with XSPEC software ver. 12.7 \citep{Arnaud1999} using one or two thermal APEC models \citep{Smith1999} 
with foreground absorption and with both fixed and varying metallicity ($Z$). 
{ We have verified that reliable spectral fits require approximately a minimum of 
500 total (MOS1 + MOS2 + PN) counts. We simulated spectra with different count statistics
and different shape to assess the validity of the 500 counts threshold.
Fainter sources are not suitable for meaningful spectral analysis because uncertainties become 
large and it is hard to distinguish between a thermal or a non thermal (e.g. a power law) spectrum.}

As a general criterion to determine the parameters from the best fit modeling, we used the $\chi^2$ statistics, 
accepting as a good fit the models with the lowest number of free parameters that realize $P(\chi^2 > \chi^2_0) > 0.01$. 
In the case of two very bright sources, we accepted the best fit parameters from 2-T and Z variable model
even if $P(\chi^2 > \chi^2_0) < 0.01$ because the overall shape of the spectrum was 
well fit. Only near few blends of lines the agreement with the best fit model was poor. 

In one case (SOXS258), we adopted a more specialized spectrum constituted by
a sum of one VAPEC + one APEC model, plus absorption. The Fe and Z abundance were linked together: the
VAPEC component takes into account a cool component while the APEC component models the high energy
part of the spectrum. Specific abundances were: [Fe/H] = 0.13 and [O/H] = 0.34, while temperatures
were found: $kT_1 = 0.2$ keV, $kT_2 = 0.5$ keV, and absorption$N_H = 1.5\cdot10^{20}$ cm$^{-2}$.
The results of the best fit procedure for the 232 bright sources are listed in Table \ref{bestfit}.
{ The rate-to-flux conversion factors (CF) for faint sources have been calculated by taking 
the median of temperatures and absorption obtained from the spectral fits of the bright sources 
for each IR class. 
With PIMMS software and by assuming an absorbed 1T APEC thermal model, the  CFs have been 
calculated for the MOS camera and the filter used given that we derived the rates for MOS with the wavelet 
detection code. The CF values are: $1.5\times10^{-11}$, $9.74\times10^{-12}$, $6.71\times10^{-12}$ 
erg cm$^{-2}$ cnt$^{-1}$ for Class I, II and III objects respectively.}

\begin{table}[t]
\caption{\label{bestfit} Best fit parameters 1$\sigma$ errors obtained from the modeling of X-ray spectra. 
We list identifier, model, value of reduced $\chi^2$, degrees of freedom, $N_H$ absorption, $kT_1$, 
normalization of the first component, $kT_2$, normalization of the second component, global abundance $Z$, 
and unabsorbed fluxes in 0.3-8 keV band.} \small
\resizebox{\columnwidth}{!}{
\begin{tabular}{llrrrrrrrrr} \hline \hline 
\soxs Id & Model & $\chi^2_{red}$ & D.o.F. & $N_\mathrm H / 10^{20}$ &  $kT1$ & $10^5\times$norm.$_1$    & $kT 2$  & $10^5\times$norm$_2$ & $Z$ & $10^{14}\times f_\mathrm X$  \\
	&  &  &   &  	  (cm$^{-2}$ )  & (keV) & (cm$^{-5}$ ) &  (keV) &  (cm$^{-5}$) &  & (\fxu)   \\\hline 
4	 & 	2t\_zvar	 & 	1.08	 & 	329	 & 	14.21$\pm$0.73	 & 	10.86$\pm$7.65	 & 	53.37$\pm$8.78	 & 	1.01$\pm$0.03	 & 	283.8$\pm$39.64	 & 	0	 & 	219.73  \\
6	 & 	1t\_zvar	 & 	1.16	 & 	8	 & 	1.94$\pm$5.04	 & 	0.86$\pm$0.09	 & 	9.58$\pm$6.17	 & 	--	 & 	--	 & 	--	 & 	5.86  \\
10	 & 	1t\_zvar	 & 	1.44	 & 	38	 & 	8.17$\pm$2.28	 & 	0.76$\pm$0.07	 & 	24.53$\pm$9.61	 & 	--	 & 	--	 & 	--	 & 	11.31  \\
11	 & 	1t	 & 	0.95	 & 	44	 & 	96.54$\pm$22.34	 & 	17.13$\pm$13.5	 & 	9.83$\pm$1.61	 & 	--	 & 	--	 & 	--	 & 	16.12  \\
16	 & 	1t\_zvar	 & 	1.45	 & 	34	 & 	17.35$\pm$3.93	 & 	0.95$\pm$0.13	 & 	19.62$\pm$10.63	 & 	--	 & 	--	 & 	--	 & 	11.51  \\
18	 & 	2t\_zvar	 & 	1.65	 & 	21	 & 	10.88$\pm$8.1	 & 	0.75$\pm$0.12	 & 	5.8$\pm$5.08	 & 	0.22$\pm$0.06	 & 	10.6$\pm$20.88	 & 	0.1	 & 	3.28  \\
20	 & 	1t\_zvar	 & 	1.01	 & 	110	 & 	4.12$\pm$1.13	 & 	0.95$\pm$0.04	 & 	24.09$\pm$4.56	 & 	--	 & 	--	 & 	--	 & 	14.46  \\
27	 & 	1t	 & 	1.56	 & 	39	 & 	0	 & 	1.54$\pm$0.09	 & 	3.99$\pm$0.37	 & 	--	 & 	--	 & 	--	 & 	3.96  \\
32	 & 	1t\_zvar	 & 	1.24	 & 	33	 & 	2.51$\pm$2.2	 & 	0.92$\pm$0.07	 & 	10.37$\pm$3.83	 & 	--	 & 	--	 & 	--	 & 	6.32  \\
33	 & 	1t	 & 	1.52	 & 	34	 & 	0	 & 	0.94$\pm$0.05	 & 	1.41$\pm$0.19	 & 	--	 & 	--	 & 	--	 & 	1.49  \\
\hline
\end{tabular}
} %
\\
\footnotesize{First 10 rows are shown, full table is published in electronic format only.}
\end{table}

For the northern fields  $N1$, $N2$, $N3$, 
we have followed the same procedure of filtering and source detection as used for the southern
fields. The total number of detected sources in fields $N1$, $N2$ and $N3$ is 864. 
From source count rates and rate-to-flux conversion factors derived for the
L1641 sources, we have derived fluxes and luminosities in 0.3--8.0 keV band. 
\begin{table}[t]
\caption{\label{tabIR} List of X-ray and IR matches (first 10 rows). 
Columns are: SOXS source number, IR coordinates, 2MASS J, H, K magnitudes, 
IRAC [3.6], [4.5], [5.8], [8.0], MIPS [24] band magnitudes, 
a flag indicating Class 0/I/flat-spectrum object, Class II object, or Class III candidate status, 
and the match radius between X-ray and IR positions. {X-ray sources with
more than a potential IR counterpart have multiple entries.}
}\small
\resizebox{\columnwidth}{!}{
\begin{tabular}{lrr|rrrrrrrr|rrr|r}\hline \hline \small
SOXS ID	 & 	R.A.	 &	Dec.	 & 	$J$	 & 	$H$	 &	$K$	 &	[3.6]	 &	[4.5]	 &	[5.8]	 &	[8.0]	 &	[24]	 & 	0/I/flat & II 	 & III & $r$ \\\hline
	 &deg (J2000)	 & deg (J2000)	 & \multicolumn{8}{c|}{mag}											 & 		 & 		 &	 	 & \arcsec \\\hline
1	 & 	83.5656	 & 	-6.6012	 & 	10.59	 & 	9.65	 & 	8.97	 & 	7.64	 & 	7.22	 & 	6.85	 & 	6.09	 & 	2.79	 & 	 F 	 & 	 T 	 & 	 F 	 & 	1.6  \\
2	 & 	83.7179	 & 	-6.5857	 & 	13.28	 & 	12.74	 & 	12.43	 & 	12.05	 & 	11.96	 & 	11.86	 & 	11.91	 & 	 -- 	 & 	 F 	 & 	 F 	 & 	 T 	 & 	1  \\
3	 & 	83.7947	 & 	-6.579	 & 	10.41	 & 	10.19	 & 	10.15	 & 	10.14	 & 	10.12	 & 	10.12	 & 	10.1	 & 	 -- 	 & 	 F 	 & 	 F 	 & 	 T 	 & 	4.2  \\
4	 & 	83.6703	 & 	-6.5764	 & 	10.68	 & 	10.01	 & 	9.84	 & 	9.76	 & 	9.75	 & 	9.7	 & 	9.64	 & 	 -- 	 & 	 F 	 & 	 F 	 & 	 T 	 & 	1.2  \\
5	 & 	83.6488	 & 	-6.5748	 & 	15.33	 & 	14.65	 & 	14.32	 & 	13.97	 & 	13.85	 & 	13.91	 & 	13.74	 & 	 -- 	 & 	 F 	 & 	 F 	 & 	 T 	 & 	2.2  \\
6	 & 	83.7836	 & 	-6.5635	 & 	8.81	 & 	8.8	 & 	8.74	 & 	8.77	 & 	8.78	 & 	8.75	 & 	8.74	 & 	 -- 	 & 	 F 	 & 	 F 	 & 	 T 	 & 	1.2  \\
7	 & 	83.7815	 & 	-6.5608	 & 	13.17	 & 	12.44	 & 	12.3	 & 	12.12	 & 	12.07	 & 	12.05	 & 	12.13	 & 	 -- 	 & 	 F 	 & 	 F 	 & 	 T 	 & 	3.3  \\
8	 & 	83.7955	 & 	-6.5514	 & 	14.39	 & 	13.75	 & 	13.42	 & 	13.21	 & 	13.11	 & 	13.11	 & 	12.93	 & 	 -- 	 & 	 F 	 & 	 F 	 & 	 T 	 & 	3  \\
9	 & 	83.7067	 & 	-6.5498	 & 	14.11	 & 	13.54	 & 	13.25	 & 	12.87	 & 	12.77	 & 	12.71	 & 	12.66	 & 	 -- 	 & 	 F 	 & 	 F 	 & 	 T 	 & 	0.6  \\
10	 & 	83.8235	 & 	-6.54	 & 	12.05	 & 	11.42	 & 	11.19	 & 	11.05	 & 	10.96	 & 	10.91	 & 	10.81	 & 	 -- 	 & 	 F 	 & 	 F 	 & 	 T 	 & 	1.3  \\
\\ \hline
\end{tabular}
} %
\footnotesize{First 10 rows are shown, full table is published in electronic format only.}
\end{table}

\subsection{\spitzer\ data.\label{spitzer_classification}}
The analysis of the \spitzer\ data and the IR catalog are discussed in \citet{Megeath2012}. 
We used the IR catalog of Megeath et al. of objects detected in the four 
IRAC bands and MIPS [24] band. 
{ The PSF width of \spitzer\ IRAC is $\sim 1.6\arcsec$ and thus better than the \xmm\ PSF, 
while the PSF width of MIPS [24] is slightly worse ($d\sim6\arcsec$).}
The \xmm\ fields are almost fully contained in the \spitzer\ sky coverage, with exception of small
portions of fields $S2$, $S3$, $S8$ and $S9$, where a total of 16 X-ray sources lie outside the 
field of view of \spitzer.

\begin{figure*}
\begin{center}
\includegraphics[width=0.48\columnwidth]{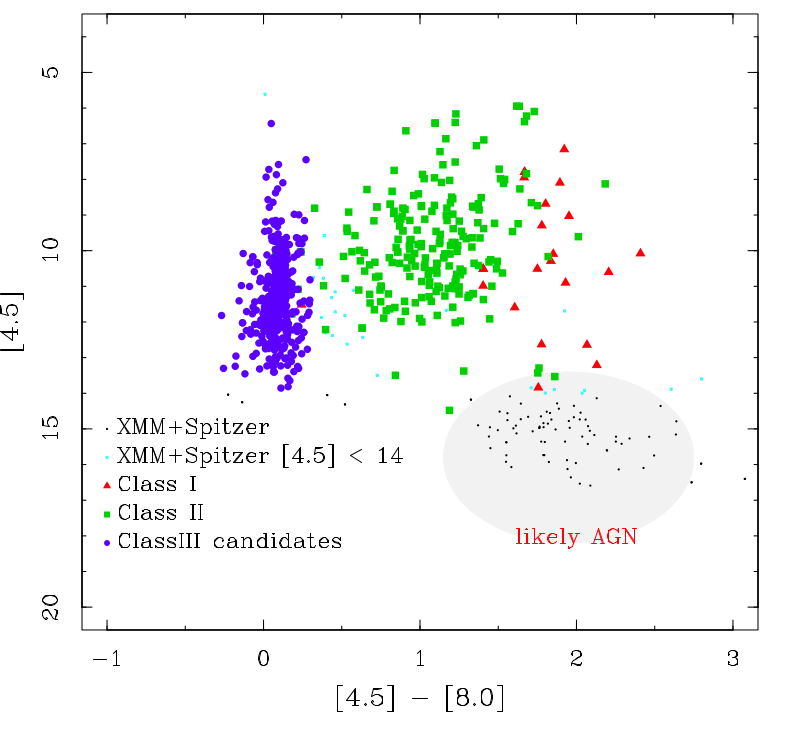}
\includegraphics[width=0.48\columnwidth]{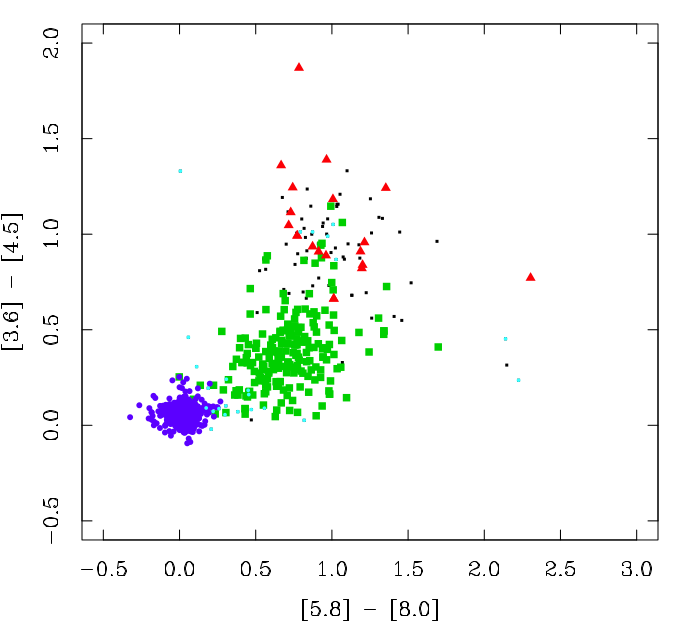}
\includegraphics[width=0.48\columnwidth]{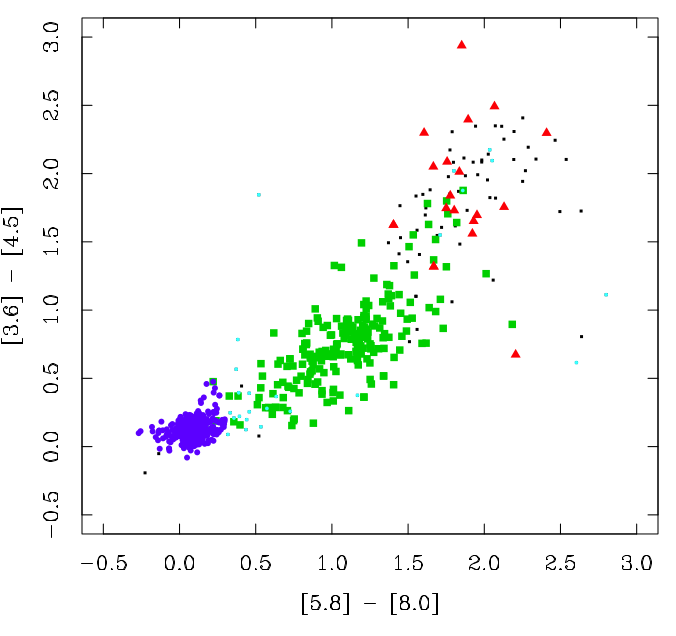}
\includegraphics[width=0.48\columnwidth]{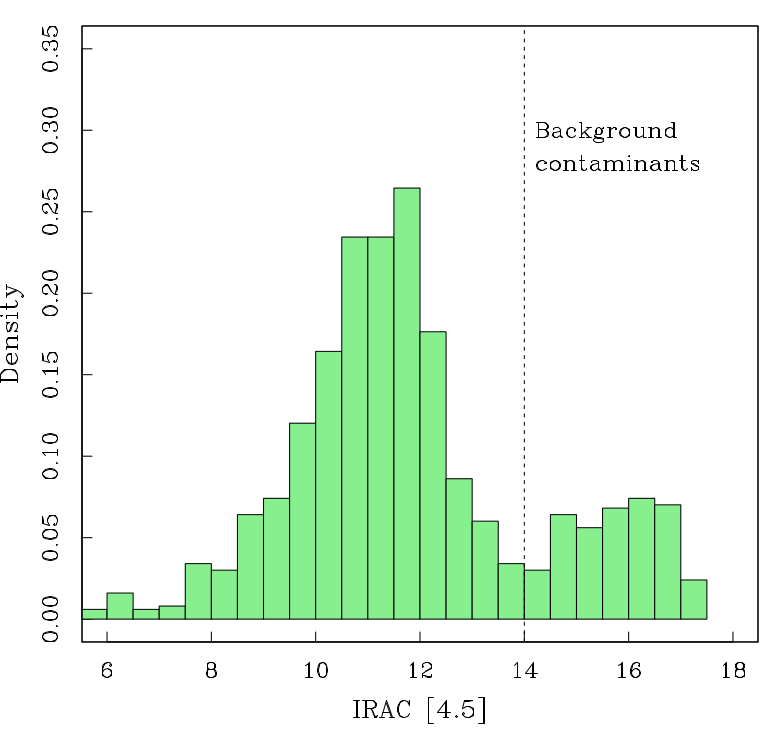}
\end{center}
\caption{\label{plots_irac} IRAC color-color and color-magnitude diagrams of \soxs objects. 
Triangles are protostars, squares are Class~II stars, as derived from \spitzer\ data alone. 
Circles are Class III  candidates. The histogram of IRAC [4.5] magnitudes shows a bi-modal population.
The objects in the fainter peak with IRAC [4.5]  $> 14$ mag are likely background objects.}
\end{figure*}

The stellar objects in the catalog have been separated into infrared ``Classes" 
(see, e.g. \citealp{Lada1992}; \citealp{Allen04}; \citealp{Megeath04}) 
following a prescription similar to that used by \citet{Gutermuth2008,Gutermuth09,Kryukova2012,Megeath2012}. 
The classification is based on cuts in a multidimensional color space, including J, H, K$_S$, 
all four IRAC channels and the MIPS 24 $\mu$m band. 
Examples of cuts in the multidimensional 
color space are shown in Fig. \ref{plots_irac}. Color-magnitude diagrams (CMDs), 
especially [4.5] versus [4.5]-[8.0] (left top panel in Fig. \ref{plots_irac}) 
help to identify the type of objects more reliably since AGNs, while red, 
tend to be fainter than young stellar objects. 
Sources are classified as either pre-main sequence stars with disks or protostars based primarily
on the slope of their spectral energy distribution, 
as described in \citet{Kryukova2012} and \citet{Megeath2012}.
Red objects fainter than [4.5] = 14 mag are usually found to be galaxies or background stars. In fact, a 
histogram of [4.5] magnitudes (Fig. \ref{plots_irac}) shows a double peaked 
distribution with a minimum between the two peaks at $[4.5] = 14$ mag. 
The main peak is composed of objects likely belonging entirely to the cloud, the second peak 
is formed by distant objects, including background stars and galaxies.

In the \xmm\ fields $S1-S10$ there are 650 Class II objects (named {\em stars with disks} in 
\citealp{Megeath2012}) and 147 Class 0-I and flat-spectrum objects ({\em protostars} in \citealp{Megeath2012}).
The total \spitzer\ survey of Orion A and B clouds is comprised of 3480 YSOs, 2992 Class II 
objects and 428 protostars. To these should be added other 50 protostars candidates and 10 sources 
detected only in \spitzer-MIPS which are likely protostars \citep{Megeath2012}.
X-rays can effectively inform us about the Class III / wTTs population of L1641. 
We have identified Class III candidates as those objects with X-ray detection,
and IRAC [4.5] $\leq 14$ mag and IRAC color $[4.5]-[8.0] \leq$ 0.5 mag (see Fig. \ref{plots_irac}). 
 
Table \ref{tabIR} lists the matched between X-rays and IR sources.

\section{Results} \label{results} 
Fig. \ref{soxsmap} and \ref{northernfields} (left panels) show an RGB image of the entire survey obtained with 
EPIC {\em XMM-Newton}. We have labeled the fields as in Table \ref{fields}. Sources which are red to 
yellow have softer, less absorbed spectra than green or blue sources. The right panels in both figures 
show the $A_V$ extinction map, adapted from Fig. 4 of \citet{Gutermuth2011}, and the X-ray sources detected in \xmm\
images. The X-ray sources show some degree of clustering where the extinction is higher, for example in fields $S5$, $S6$ and $S10$.
We detected a total of \srcnr X-ray sources in the ten southern \soxs fields ($S1$ to $S10$) with more than 
$4.5\sigma$ significance. 
The 1060 sources include 31 sources detected in the overlapping regions 
between two adjacent pointings. 
We match the X-ray positions with positions of near- and mid-IR sources from
the \spitzer\ catalog, at an offset limit threshold of 5$\arcsec$ and iterated the cross identification
process three times to remove systematic offsets along R.A. and Dec of the X-ray sources. 
{ The final offsets for sources with off-axis less than $10\arcmin$ from the aim point 
have a median of $1.4\arcsec$, and a $10\%-90\%$ range of $0.7\arcsec - 3.3\arcsec$. 
For sources at off-axis $>10\arcmin$, the values are similar: the median of offsets is $1.7\arcsec$, and
 $0.7\arcsec - 3.7\arcsec$ is the $10\%-90\%$ range.}

\begin{table} \small
\caption{\label{transdisk} 
List of candidates with MIPS [24] excesses, indicating transitional disks and two additional 
Class III candidates, which are IR objects with incomplete \spitzer photometry and a match with X-ray sources.}
\begin{tabular}{c rr|rrrrrrrr|l}\hline \hline
Id  & R.A.& Dec. (J2000, deg)  & J     & H     & K$_s$ & [3.6] & [4.5]& [5.8] & [8.0] & [24]  & Notes \\ 
    & (J2000, deg) & (J2000, deg) & \multicolumn{7}{c}{(mag)}      &                       & \\ \hline
 13 &  83.525907 & -6.5125008 & 11.59 & 10.84 & 10.59 & 10.42 & 10.39 & 10.34 & 10.30 &  5.68 & $^a$, $^b$ \\
 33 &  83.905423 & -6.3907431 & 12.52 & 11.87 & 11.61 & 11.40 & 11.32 & 11.27 & 11.19 &  5.94 & $^a$ \\
136 &  84.385151 & -6.6514246 & 12.97 & 12.28 & 12.07 & 11.83 & 11.77 & 11.71 & 11.59 &  6.88 & $^a$ \\
163 &  85.734935 & -8.6293451 & 13.74 & 12.87 & 12.57 & 12.33 & 12.27 & 12.19 & 12.12 &  9.14 & $^c$ \\
173 &  85.269234 & -7.7611424 & 11.63 & 10.67 & 10.30 & 10.11 & 10.04 & 9.916 &  9.80 &  6.17 & $^a$ \\
269 &  85.475248 & -7.8314385 & 12.23 & 11.45 & 11.08 & 10.80 & 10.72 & 10.67 & 10.59 &  6.52 & $^a$ \\
276 &  84.907365 & -7.4397500 & 16.23 & 13.50 & 12.12 & 11.34 & 11.08 & 10.86 & 10.86 &  7.48 & $^a$ \\
281 &  84.640908 & -7.2841951 & 11.09 & 10.36 & 10.14 & 10.01 &  9.99 &  9.96 &  9.91 &  7.34 & $^a$ \\
383 &  83.862722 & -5.8145652 & 10.48 &  9.99 &  9.80 & 9.035 &  8.74 &  8.37 &  7.73 &  4.68 & $^a$ \\
468 &  83.946119 & -6.1958249 & 13.86 & 13.25 & 12.97 & 12.59 & 12.48 & 12.39 & 12.36 &  8.86 & $^d$ \\ \hline
 61 &  83.820037 & -6.3040907 & --    & --    &  --   & --    & --    & 11.05 &  --   &  --   & $^e$ \\
126 &  84.510766 & -6.7102342 & --    & --    &  --   & 14.77 & 14.68 & --    &  --   &  --   & $^f$ \\\hline
\end{tabular}
Notes: $^a$ Strong, isolated detection. $^b$:  off of the main cloud region. 
$^c$: faint, near airy ring. $^d$: faint, in nebulosity, near the edge of 
detection, maybe elongated, it could be a contamination. $^e$: double source. $^f$: faint, only detected in IRAC
[5.8] and [8.0].
\end{table}

We find that 972 X-ray sources have at least a  match with an object in the 2MASS$+$\spitzer\ catalog.
There are 69 cases of X-ray sources with multiple counterparts among the IR objects: 66 cases 
of matches with two IR objects and 3 cases of triple matches.

We have also considered X-ray sources with a significance of detection lower than the 4.5$\sigma$ threshold, 
in order to identify other faint X-ray emission among known Class I and Class II objects. 
With this choice, we find nine protostars and 19 Class II objects with X-ray detection, which have 
a significance  between 3 and 4.5 $\sigma$ of the local background. 
On average, one additional Class I and two Class II objects per field are found, respectively. 
The probability of a chance match of these stars is of order of $10^{-3}$. 
However, in the following analysis we will not consider these additional X-ray detections 
and we will only discuss the
sample of 1060 sources detected with significance above $4.5\sigma$.
In particular, we cannot extend the list of Class III candidates by considering low significance detections, 
because of the high number of objects in the IR catalog which appear as normal photospheres and, 
consequently, the high likelihood of matching spurious sources.
Table \ref{detlist} reports the number of \spitzer\ objects in fields $S1-S10$, 
the number of X-ray sources, and the rate of detections from the list of 1060 X-ray sources.
We give also the ratios of Class I to Class II objects and of Class III to Class II objects.
Analogous numbers for the fields $N1-N3$ are biased and incomplete because of source confusion in 
X-rays and IR bands around the core of the ONC.
{ By following the criteria in \S \ref{spitzer_classification}, we find 489 Class III candidates. 
Among these 489 objects, ten objects have a strong MIPS 24  $\mu$m band excesses that suggest they are
transitional disks objects, i.e. young stars with a depleted inner circumstellar disk. 
Two further objects have incomplete photometry in IRAC and MIPS bands that prevents the detection
of IR excesses. One objects is a very faint IR source with detection only in [5.8] and [8.0] bands, 
the other object is actually a double source with confused photometry. 
The positions and the IR photometry of the transitional disk objects and the two faint objects 
are shown in Table \ref{transdisk}. }

We give an estimate of the number of extra-galactic sources that we expect to find by using PIMMS 
and the limiting count rate given in \S \ref{sensitivity}.
To estimate a limiting flux for these objects, we assumed a spectrum due to an absorbed 
power law model with index 1.4 for the case of self-absorbed AGNs and 1.9 for not self-absorbed AGNs. 
Then we derived the number of expected extra-galactic sources per square degree by using 
the $\log N \log S$ model of {\em POMPA} code (\citealp{Gilli07}; see also \citealp{Puccetti09}). 
The model predicts $\sim70-90$ sources per square degree in $0.5-10$ keV or about 120-150 sources in SOXS.  
{ Fig. \ref{plots_irac} shows about 75 objects detected in X-rays and with IRAC colors and magnitudes 
consistent with those of background AGNs. }
This number is lower by a factor $\sim1.6-2$ with respect to the number of expected extra-galactic sources. 
A possible bias of this estimate is the presence of the cloud of Orion A and its effect on X-ray fluxes.
 
{ Nineteen X-ray sources are unidentified in IR or optical catalogs}. 
{Statistically, about ten to fifteen spurious sources are expected for the adopted detection threshold, 
as estimated from the simulations of background only images.
We expect that the spurious sources are among those with lowest significance.}
We have thus an excess of about five to ten unidentified {\em true sources}. 
Given they are not matched in the IR catalog, their nature could be extra-galactic or
they are still YSOs embedded in bright nebulosity that hampers the IR detection.  

\begin{table*}[!t]
\caption{\label{det_table} Number of \spitzer\ objects, X-ray detections, Class I and Class II YSOs 
in each \soxs field.}
\begin{center}\small
\resizebox{\columnwidth}{!}{
\begin{tabular}{l c c c c c c c c c c | r }\\ \hline \hline
Field name            & S1	& S2	& S3 	& S4	&S5a+b& S6	&S7	& S8	& S9	& S10	& Total\\ \hline
N$_\mathrm{Spitzer}$  &  4232	& 4596  & 4641	&  4453 & 3341	&4345	& 4259	& 4135	& 4259	& 4477	& 39891	$^a$  \\
Class I               & 12	&  4	&  13	&  8	&  12	&  31	&  6	& 35 	&  6	&  16	&  147 	\\ 
Class II              & 69	& 36	& 110	& 39	&  147	&  54	& 62	& 55 	& 62	&  43	&  650 \\
Cl I / Cl II          & 17\%	& 11\%	& 12\%	& 21\%	&  8\%	& 57\%	& 10\%	& 64\%	& 10\%	& 37\%	&  23\% \\ \hline
N$_X$                 & 202	& 149	& 161	& 104	&  83	& 69	& 53	& 84	& 107	& 48	& 1060 \\
N$_\mathrm{X}$ in \spitzer& 200	& 156	& 158	& 104	&  76	& 66	& 51	& 78	& 96	& 56	& 1041 \\
X det. Class I        & 1	&  2	&  4	&  5	&  1	& 3	& 1	& 1	& 1	& 2	&  23 \\
X det. Class II       &  19	& 31	& 44	& 15	& 10	& 16	&  9	&  12	&  9	& 21	&  204 \\ 
X det. Class III cand.&  107	& 113	& 78	& 15	& 50	& 14	& 23	&  20	& 49	& 27	& 489 \\ 	
\hline
Det. Fract. Cl I      & 0.08	&0.50	& 0.31  & 0.63  & 0.08	& 0.1	& 0.12  &  0.03	&  0.17  & 0.63  & 0.14$^b$  \\
Det. Fract. Cl II     & 0.27	&0.86	& 0.40  &  0.39 & 0.07  &0.30	& 0.49  & 0.22	&  0.15  &  0.39 & 0.29$^b$  \\
X det. Cl I / Cl II   & 0.09	&0.33 	& 0.06	&  0.10 & 0.05  &0.19	& 0.11  &  0.08	&  0.11  &  0.10 & 0.11 \\  
X det. Cl III/Cl II   & 5.60	&3.65	& 1.77	& 1.00	& 5.00	&0.87	& 1.28  & 1.67	&  5.44	 & 1.00	 & 2.28$^b$ \\
\hline
\end{tabular}
} %
\end{center}
Note: $^a$ Some of the \spitzer\ objects are in overlapping fields, so that the sum of the numbers in each field 
is greater than the total of \spitzer\ objects present in all fields.
$^b$ This is the average detection rate, i.e. the sum of detections divided by the sum of objects in each field.
\end{table*}

\subsection{Sensitivity of the survey \label{sensitivity}}
We have calculated sensitivity maps for each field based on the count rate threshold 
used for the source detection. All southern fields ($S1-S10$), except those pointed toward V883 Ori 
($S5a+b$) and $\iota$ Ori ($S1$), have similar exposure times and similar 
instrumental sensitivity. 
Globally, the limiting sensitivity for a point source is $\sim 1.8\times 10^{-4}$ ct s$^{-1}$
at the center and it is reduced to $\sim 10^{-3}$ ct s$^{-1}$ at $\sim13^\prime$ off-axis. 
The combined V883 fields ($S5a+S5b$) are shallower and have about half the sensitivity of other fields. 
The same is true for field $S1$.
Field $N1$ is similar to $S5a+b$, while $N2$ is about twice as deep as the rest. 

To translate the rate sensitivity into limiting flux, we have considered a mono-thermal absorbed APEC model. 
The absorption column density is the main factor affecting the limiting flux, overwhelming the 
effects of a change in the plasma temperature From spectral analysis, we have derived values of 
absorption as high as $N_H \simeq 5\times10^{22}$ cm$^{-2}$ with a median value of 
$N_H \simeq 7\times10^{20}$ cm$^{-2}$.
Plasma temperature are found in the range between $kT= 0.7$ keV and $kT=3$ keV. 
We used the PIMMS software to estimate the limiting flux to the typical coronal emission
of young stars. For a plasma with a temperature of $kT = 1 $keV and for an absorbing gas column 
$N_\mathrm H$ ranging from $10^{19}$ cm$^{-2}$  to $5\times10^{22}$ cm$^{-2}$ (the range of values
that we derived from the best fit models), we have a limiting unabsorbed flux in the range of
\fx $= 1.4\times10^{-14}-6.6\times10^{-13}$ \fxu {($0.3-8.0$ keV)}, which is a variation of 
a factor $\sim47$. 
By considering a hotter plasma with $kT= 3$ keV, and varying again 
$N_H$ from $10^{19}$ cm$^{-2}$ to $5\times10^{22}$ cm$^{-2}$, the limiting unabsorbed flux varies by a factor 4.
The effect of absorption on the spectrum is less pronounced on the spectrum with a hotter kT because 
most of the emission is above 1 keV where absorption is not relevant. 

The limiting flux is determined mainly by the range of $N_\mathrm H$ absorption present in our survey, rather 
than the effect due to a hotter or cooler plasma. 
Given the range of unabsorbed fluxes given above, at the distance of the ONC (414 pc, \citealp{Menten07}), 
our survey can detect any source with an unabsorbed X-ray luminosity 
of $\log$ \lx $ \geq 28.5$ for a source placed 
on-axis. EPIC camera has a quite stable PSF width within $\sim10\arcmin$. 
For sources placed off-axis by more than 10$\arcmin$, the decrease of sensitivity is about a factor $\sim2$.

\subsection{Completeness and contamination.}
The completeness of the IR catalog by \citep{Megeath2012} depends on the luminosities of the stars, 
the presence of a disk, and the brightness of the surrounding nebulosity.  
About 90\% completeness is obtained at $K = 13$ mag. By using the tracks of \citet{Baraffe1998},
this translates into 0.09 $M_\odot$ at 1~Myr and 0.175 $M_\odot$ at 3~Myr, equivalent to a $\sim$M4 type
\citep{Luhman1999,Kenyon1995}. 
In dense regions like the ONC the limit at the $K$ band can be two to three magnitudes brighter.
In regions of low nebulosity like L1641 this effect is only limited to the densest parts and
with an increase of the limiting magnitude of about 1 mag like in L1641S across fields $S8-S9$. 
As a consequence, the completeness of the survey in denser regions is decreased to $\sim0.175 M\odot$ 
for an age of 1M~yr and 0.35 $M\odot$ at 3~Myr.

We assess the  completeness of PMS stars detected in X-rays by comparing
to the results from COUP \citep{Feigelson05,Getman05}, which constitutes the most complete sample 
obtained in X-rays for a star forming region ($\sim 97\%$ complete above spectral type M4). 
The XLD is modeled with a log-normal distribution centered at log \lx\ $= 29.3$ and standard 
deviation equal to 1 dex. If we assume that this XLD yields also for TTs in other SFRs, 
we conclude that our X-ray observations have probed about 45\% of the total population of TTs.
{\citet{Getman2006} have modeled the XLD of lightly obscured YSOs in COUP
by means of a broken power law, with a knee at $\log L_X = 2.5\times 10^{30}$ erg/s, while in
other star forming regions, like NGC~1333 and IC~348, a log-normal model still fits the data.
By adopting the broken power-law model as in \citet{Getman2006}, we obtain a completeness at a level of 46\%.}
The X-ray sample is complete to $\log L_X \sim 29.5$, which corresponds approximately to 
the luminosity of an early K-type star at age of 3 Myr.

The number of expected foreground contaminants in X-rays has been estimated 
by means of a log-normal distribution of the X-ray luminosities of field stars 
(XLF$_{field}$) with median $\log L_X = 27.5$ 
and standard deviation $\sigma = 1$ dex \citep[see][and references therein]{Favatarev03}. 
We have assumed a conservative limiting flux of $f_{X,lim} = 1.5\times 10^{-15}$ 
\fxu\ and integrated all the contributions of the XLD$_{field}$  up to the distance to Orion 
by assuming a density of field stars of 0.1 pc$^{-3}$. At most
five foreground stars per field, $\simeq50$ stars in all fields are expected to be detected 
in {\em SOXS}. \label{complete}

\begin{figure}
\begin{center}
\includegraphics[width=0.79\columnwidth,angle=0]{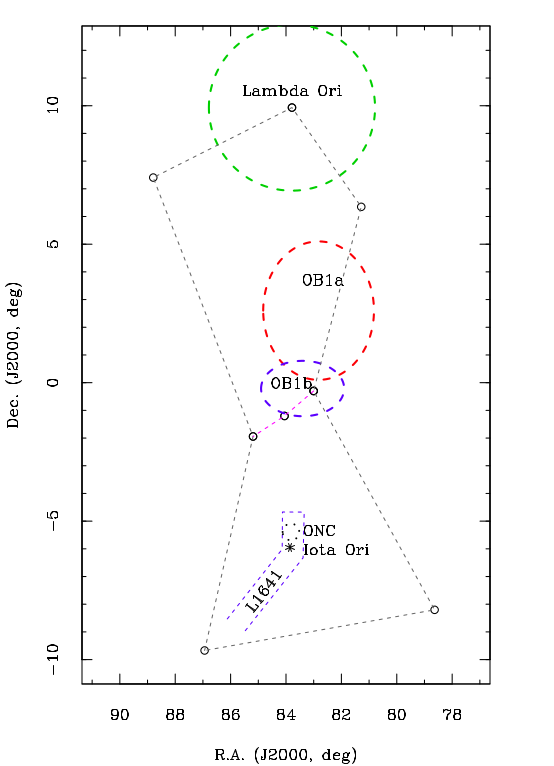}
\end{center}
\caption{\label{sim_scheme} Geometry adopted for Monte Carlo simulations of contaminants from 
OB1a, OB1b, and the ONC. Locations of subgroups are taken from \citet{Bally08}. 
The region around $\lambda$ Ori has been taken into account but given the distance
and the age it gives negligible contribution to the PMS population in L1641. The
same is true for OB1b stars. }
\end{figure}

\begin{table}
\caption{\label{cont-tab} Expected number of YSOs from Orion OB1a and the ONC
in the region of the survey.}
\begin{center}
\begin{tabular}{ l  c c }\\\hline \hline
Field	&	From OB1a&	From the ONC \\ \hline
S1	&	2.6	 & 	104.9	 \\
S2	&	2.1	 & 	16.7	 \\
S3	&	2	 & 	13.7	 \\
S4	&	1.6	 & 	1	 \\
S5	&	1.3	 & 	0	 \\
S6	&	1.1	 & 	0	 \\
S7	&	0.9	 & 	0	 \\
S8	&	0.7	 & 	0	 \\
S9	&	0.5	 & 	0	 \\
S10	&	0.4	 & 	0	 \\\hline
Total	&	13.1	 & 	136.3	 \\\hline
\end{tabular}
\end{center}
\end{table}

We have also considered the possibility of detecting stars born in the northern part of Orion, 
namely the Orion OB1a and OB1b star forming regions, that could have traveled south 
since their birth to the Orion~A cloud and L1641. 
To estimate the number of these objects, we considered a normal distribution of stellar velocities 
with zero mean and dispersion equal to 3 km s$^{-1}$, and with an age of 10 Myr, 3 Myr, and 1 Myr  
for OB1a, OB1b and the ONC, respectively (see \citealp{Bally08}, \citealp{Buckle2012}). We make the
assumption that the group of stars from the northern regions is, on average, co-moving with L1641.
The stars with velocities of 3 km s$^{-1}$ should have traveled
on average distances of 30 pc, 9 pc and 3 pc from their birth place,  
which correspond to an angular distances of  $5.32\deg$, $1.18\deg$ and
$0.42\deg$, taking into account the distances of these groups (OB1a 330 pc, OB1b 440 pc 
-- \citealp{Sherry04}; and the ONC 414 pc). 
Furthermore, we have assumed a population of 1,000 stars for OB1a, OB1b and the ONC, 
respectively. We ran 10,000 Monte Carlo simulations of the stellar populations with ages as above 
with a normal distribution of velocities with mean of zero and standard deviation of 3 km s$^{-1}$.
The results are shown in Table \ref{cont-tab} where we report the expected contaminants for each \xmm\ field.
The number of contaminants from OB1a decreases North to South with a total of $\sim13\pm3$ stars. 
The contamination from OB1b and the $\lambda$ Ori ring are negligible due to the
combined factors of distance and age, and thus their values are not reported in Table \ref{cont-tab}.
About eight stars may have formed in OB1a and then traveled into 
NGC~1980 and L1641~N (fields S1+S2+S3+S4).

The contamination from the ONC is significant in the northern fields of {\em SOXS}. 
As shown in Table \ref{cont-tab}, the total number of objects traveling from the ONC to NGC1980 and 
L1641~North is $\sim136\pm12$ stars.
Due to the difference in ages of the ONC and OB1c to which NGC1980 belongs (1 Myr vs. $\sim5$ Myr, \citealp{Bally08}), 
we expect that the population of YSOs coming from the ONC should be less evolved than the original population
of stars in OB1c, and thus preferentially of earlier Classes (0/I, and II).

In summary, we estimate that less than fifteen high velocity stars from the subgroups north of the 
Orion's Belt are present in \soxs fields. 
More pronounced is the number of stars that could have born in the ONC and traveled into 
L1641~N and NGC1980, and this number is about 130 YSOs. Contamination from the $\lambda$ Ori group and
OB1b is negligible.

\section{The population of young stellar objects}\label{ysopop} 
In this section we focus on the properties of the YSOs selected by means of the
spectral energy distribution (SED) shape in \spitzer\ bands for Class I and Class II objects, 
and by their  X-ray detection and absence of IR excess for Class III candidates. 

The overall ratio of Class I to Class II objects detected by \spitzer\ is $0.23$ 
(range\footnote{Obtained from the square root of numbers in Table \ref{det_table}.}: $0.20-0.25$), 
but from field to field this ratio varies between $0.08$ ($0.05-0.11$) and 0.64 ($0.47-0.86$). 
In particular, the largest number of Class I YSOs 
with respect to Class II objects is found in the field $S8$ near L1641 South.

The ratio of Class III candidates and X-ray detected Class II objects also varies across the fields.
Values of the Class III to Class II ratio range from $0.87_{0.51}^{1.48}$ to $5.63_{4.14}^{8.01}$
with a mean ratio of $2.39 (2.14-2.69)$. The number of Class III objects is always larger than the number 
of Class II objects except in the fields $S5$ and $S6$. 
The highest ratio of Class III candidates with respect to Class II YSOs is observed in two
distinct regions: the northern fields, $S3$ and $S1$, where the ratio ranges between three and five
and in $S9$, where it is more than five. This high ratio hints that a more evolved population 
of young stars is present toward $\iota$ Orionis and V380 Ori and locally in the subgroup 
belonging to L1641S as we will discuss later. 

A main goal of this survey is to estimate the number of young stars in L1641 North, L1641 South, and 
in the fields around V380, V883 and $\iota$ Ori.
While we consider the sample of Class I and II YSOs detected and classified with \spitzer\
almost complete down to the $\sim$M4 limit \citep{Megeath2012}, 
there is considerable uncertainty in the total PMS population due to the Class III stars. 
These appear as normal photospheres in the \spitzer\ bands, but
they are characterized by high X-ray luminosity, in a range from  $5\times10^{28}$ erg s$^{-1}$ to 
$10^{31}$ erg s$^{-1}$. In \S \ref{spitzer_classification}, we have defined Class III candidates 
as the \spitzer\ counterparts of X-rays sources  with IR magnitudes consistent with PMS stars 
([4.5] $< 14$ mag at the distance of Orion A, and without IR excess. 
The number of Class III candidates in fields $S1-S10$ is 489. 
If we subtract 50 expected foreground stars, as estimated in \S \ref{complete}, 
we are finding at least 439 Class III candidates. 
In \S \ref{sensitivity}, when discussing the sensitivity of the survey, 
we estimated a detection rate of X-ray sources of about 45\%. 
Taking this number as the detection rate of the Class III stars, 
we estimate a population of Class III stars of 1087 members in the case of 489 X-ray detected stars, 
and 976 stars when correcting for the 50 foreground objects. 
To these numbers we add 650 the Class II YSOs and 147 protostars present in the fields $S1-S10$, 
thus giving a total population of $\sim1700-1884$ ($\pm90$) of YSOs, 
in different evolutive stages, from protostars to disk-less stars. 

On the other hand,  one could consider that the fraction of  Class III stars detected in 
X-rays is the same of Class II objects ($\sim30\%$).  
Under this hypothesis, the total sample of Class III stars  should be about 1630$\pm75$ or  
 $\sim 1460\pm70$ when correcting for the foreground objects. 
This leads to a total of $\sim$2260 $\pm110$ PMS stars above mid M-type stars in the \xmm\ fields $S1-S10$. 
One bias of this estimate are the implicit assumptions that 
hat Class II and III objects have similar luminosities and absorption, similar ages and that all
the young stars are at the same distance.
This is not exactly true for the region around \iori\ where we find evidence of a closer and more evolved
population that is depleted of protostars and Class II objects.

In summary, the total population of PMS stars in the portion of L1641 
and around \iori\  surveyed with \soxs\ is comprised between $\sim1600$ and $\sim2350$ stars. 
The real number is more likely closer to the lower estimate given the caveat given above.
The YSOs population is comparable  to the ONC population, although it is dispersed over a more extended area
and lacks high-mass stars. 
The \xmm\ survey has covered only about 60\% of the \spitzer\ survey in L1641 (Fig. \ref{soxsmap}).
We cannot exclude that a significant fraction of Class III stars are present in the 
remaining part of L1641 filament and missed by \spitzer\ because of a lack of IR excess. 
The region of L1641 surveyed by \spitzer\ but not by \xmm\
contains 22 protostars and  104 Class II YSOs. By assuming the same ratio of Class III to Class II object found
in {\em SOXS}, about 365 $\pm20$ YSOs should be in the region without \xmm\ coverage. 
The total young population in L1641 is then estimated in the range $2000-2700$ members.
If the density of the cloud is a probe for the youth of YSOs embedded in it \citep{Gutermuth2011}, 
this part of L1641 could contain even more Class III stars relative to Class II objects because of 
its low density. Further X-ray observations in this region can effectively complete the census of YSOs.
\subsection{X-ray spectra and luminosities}
The analysis of EPIC spectra provides information about the absorption and the temperature of the
emitting plasma.
\subsubsection{Absorption.} 
The equivalent hydrogen column density, $N_H$, is a way to quantify the absorption of
X-rays along the line of sight which is due to atoms and ions.
On the other hand, optical and infrared extinction are due to dust grains which have 
different sizes and shapes.
Factors that can affect the $N_H/A_K$ ratio among star forming regions are changes in metallicity,
and a different mass ratio between gas and dust. 

\begin{figure}
\begin{center}
\includegraphics[width=0.79\columnwidth]{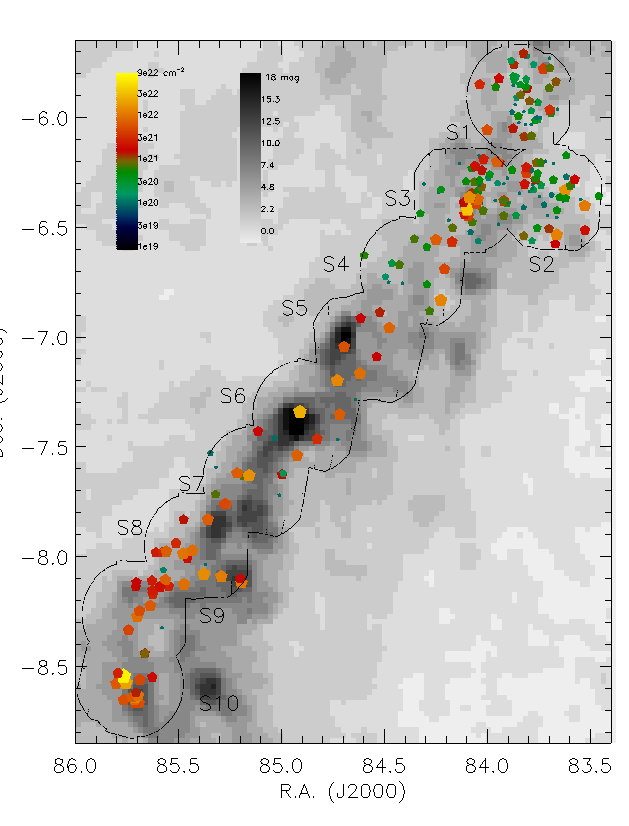}
\end{center}
\caption{\label{nh-av} $N_H$ absorption and visual extinction. 
Symbols are $N_H$ values at the positions of sources with the best-fit of the 
X-ray spectrum. Sizes and colors of the symbols code the intensity of 
$N_H$ as reported in the scale on the top-left corner of the panel.
The extinction map is adapted from Fig. 4 in \citet{Gutermuth2011}. Values of extinction
are in the top-right scale. Contours and labels refer to the \xmm\ fields as in Fig. \ref{soxsmap}. 
Highest values of $N_H$ correspond to more extincted regions. Notice the presence of low-absorption stars  
around \iori\ and L1641~N.}
\end{figure}

The  $N_H$ to $A_K$ ratio along the line of sight can be viewed in terms of gas to dust ratio, so 
it is possible to  estimate this ratio for the X-ray sources with spectral fits and
IR counterparts with a known extinction, like $A_K$. 
Our aim here is to compare the extinction $A_K$ with the absorption values of $N_H$ obtained from the 
fit to the X-ray spectra.
\label{abs} Fig. \ref{nh-av} shows a map of $N_H$ values obtained from the best fits to spectra of 
bright X-ray sources, over-plotted on the $A_V$ map (adapted from Fig. 4 of \citealp{Gutermuth2011}). 
Colors and sizes of the symbols are proportional to the $N_H$ column absorption (top-left scale). 
The $A_V$ map has been obtained from a map of $(H-K)$ values of 2MASS objects in a rectangular grid with an 
angular resolution of $\sim9^\prime\times9^\prime$. At each point of the grid, the mean of $(H-K)$ 
of the 20 nearest objects was calculated, and outliers at $>3\sigma$ were removed iteratively until 
convergence was obtained. 
From the observed $(H-K)_{obs}$ map, and by assuming a mean intrinsic color $(H-K) = 0.13$, the visual 
extinction map of  $A_V$ is obtained as: $<A_V> = 15.87 \times (<(H-K)_{obs}> - 0.1)$ \citep{Lada1994,Lombardi2001}.
As warned by \citet{Gutermuth05}, this map can be biased toward lower absorption values than the
total cloud value for the line of sight because of the presence of stars embedded in the cloud itself.

In the northern part of the map, around \iori, we observe a mix of very absorbed stars 
with $N_H>10^{21}$ cm$^{-3}$ (red and orange dots in Fig. \ref{nh-av}) and less absorbed stars 
which appear to be in front of the cloud (green symbols in the same figure). 
This is an evidence of a more evolved population of PMS stars around \iori, not embedded in the cloud 
but rather in the foreground with respect to the ONC \citep{Bally08,Alves2012}. 
The less absorbed stars in the northern part of L1641 are part of this older population
which sits in front of the cloud or at its surface where the
absorption is lower, with \iori\ being the most massive member of this foreground cluster.
On the other hand, in the southern part of L1641, strongly absorbed stars are a major fraction 
of the whole population. 
The highest values of $A_V$ are found in L1641S, where high $N_H$ 
absorption values are also measured. 

We evaluate the relationship between $N_H$ absorption and $A_K$ extinction, using the $A_K$ values from the 
individual stars themselves assuming the extinction law from \citet{Flaherty2007}.
A linear relation between $N_H$ and $A_K$ (with $A_K \sim 0.1 A_V$, \citealp{Rieke2004,Rieke1986}) yields 
values of $N_H/A_K$ in the range $1.8-2.2\times 10^{22}$ cm$^{-2}$ mag$^{-1}$ for the ISM 
\citep{Vuong03,Gorenstein75}. 
For other star formation regions (SFRs), \citet{Feigelson05} and \citet{Vuong03} have found the slopes slightly lower 
($N_H/A_K = 1.6\times 10^{22}$ cm$^{-2}$ mag$^{-1}$), 
while significantly lower values have been found by \citet{Winston2010}
for the Serpens and NGC~1333 SFRs. An intermediate slope of $1.3\times 10^{22}$ cm$^{-2}$ mag$^{-1}$ is found in 
the region of IRAS~20050+2720 \citep{Guenther2012}.

\begin{figure}
\begin{center}
\includegraphics[width=0.79\columnwidth]{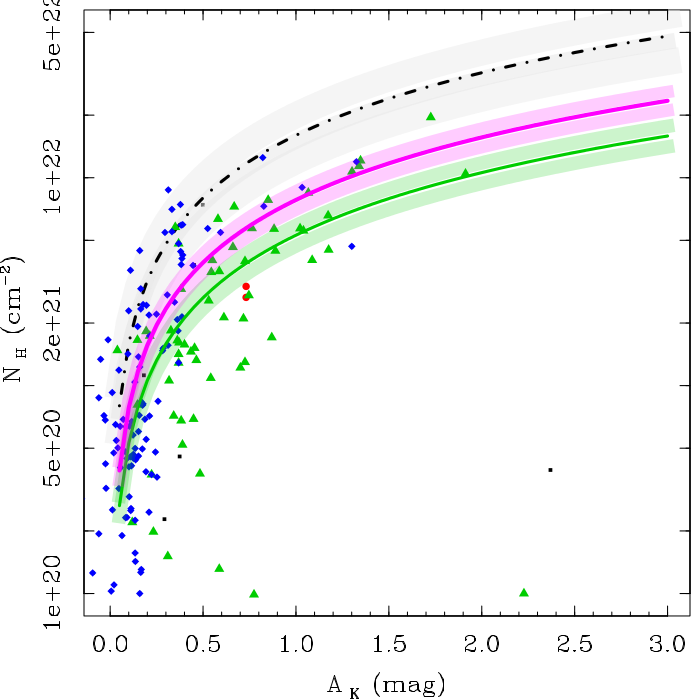}
\end{center}
\caption{Scatter plot of $N_H$ vs $A_K$ values. \label{nh-ak2}
Red circles are protostars, green triangles are Class II objects, blue diamonds
are Class III candidates. Black dots are other objects not classified as PMS stars.
Dot-dashed line is the relation derived by \citet{Vuong03}.
Solid lines are the best fits to all data (green line) 
and only objects without IR excess (magenta line): 
{ $N_H \sim (0.70\pm0.03) \times 10^{22} \cdot A_K$ and $N_H \sim (0.76\pm0.05)\times 10^{22} \cdot A_K$,
respectively. Shaded areas represent the 90\% level of uncertainty of the slope values. }
N$_H$/ A$_K$ ratio in L!641 is different than the one of the ISM but consistent with NGC~1333 and Serpens.}
\end{figure}

Fig. \ref{nh-ak2} shows the scatter plot between the column absorption values $N_H$ and $A_K$.
The linear fit of all points  gives $N_H = (7.0 \pm0.3) \times 10^{21} A_K$; by excluding 
YSOs with IR excesses we obtain $N_H = (7.6 \pm0.5) \times 10^{21} A_K$, still consistent with the 
value of the whole sample. When excluding the objects with IR excess, we reduce the influence of local 
absorption due to the presence of the disk in Class II YSOs 
and the envelope in Class I YSOs. Our best fit value is similar to those obtained by 
\citet{Winston2010} for Serpens and NGC1333 $-$ none of which contains high mass stars, and 
it is lower than those found for high mass star forming regions, such as RCW~38, RCW~108 
(\citealp{Wolk2006} and \citealp{Wolk2008}) and the ONC as well as the ISM (\citealp{Vuong03},
$N_H/A_K(ONC) = (1.0\pm0.1)\times 10^{22}$ cm$^{-2}$ mag$^{-1}$;
$N_H/A_K(ISM) \sim (2.0\pm0.2)\times 10^{22}$ cm$^{-2}$ mag$^{-1}$ ).

If we consider the $N_H/A_K$ ratio as a measure of the gas-to-dust ratio, IRAS 20050+2720, 
Serpens and NGC 1333 are dust rich (or gas poor), compared with the interstellar medium. 
In contrast, the sample from \citet{Gunther2008} shows most objects to be dust depleted (or gas
enriched) with much lower absolute values for the extinction. These sources are mostly
from the Taurus-Auriga molecular cloud complex, which is more dispersed and older than
IRAS~20050+2720, Serpens and NGC 1333, and the SFR contributes little to the line-of-sight
extinction. The extinction in Serpens, IRAS~20050+2720, and NGC~1333  
should be dominated by local circumstellar material, which
could be dust-depleted due to the energy input from the star, gravitational settling of
grains in the disk-midplane and planet formation. However, in the younger star forming
regions, we mostly see the cloud material, which could be dust-rich for those clouds without
high-mass stars, or metal poor in the gas phase due to grain formation. 

While gas depletion is a possibility, it is not the only possible cause of the anomalously low $N_H$ to 
$A_K$ ratio. Current models predict that volatiles accrete on grains in cold dense clouds, 
and the grains start sticking together so the growth occurs with time  \citep{Ormel2011}.  
Evidence for grain growth in molecular clouds has been found in the changing IR extinction law
toward high extinction line of sight \citep{Flaherty2007,Chapman2009, Mcclure2009,Chiar2007}.
{ Recent sub-millimeter observations made with {\em Herschel} by \citet{Roy2013} provide evidence 
for a change in the dust opacity which is explained as the result of the evolution of dust grains.}  

In contrast, without the condensation of volatiles, the grains reach an equilibrium size quickly 
by balancing aggregation and fragmentation.
Thus, we would expect the $N_H$/$A_K$ ratio to decrease with time in cold clouds, until the clouds are 
heated to the point that the volatiles evaporate. 
The survey of L1641 shows that the $N_H/A_K$ ratio is changing within the Orion A cloud, 
with the ONC shows an $N_H/A_K$ more typical of the ISM and other high mass star forming regions, 
while the L1641 region showing a ratio similar to that of dense molecular clouds forming clusters of 
low to intermediate stars, such as Serpens and NGC 1333 \citep{Winston2010}.  
We hypothesize that the difference between the massive and low to intermediate mass star forming 
regions is the heating of the dust by the young stars.  In the clusters of low to intermediate mass stars, 
the inner region of the clouds is shielded from the interstellar radiation field and there is relatively 
little heating from the embedded stars.  In contrast, toward massive star forming regions, the intense 
radiation fields from the high mass stars heats the dust.  
The heating of the dust raises the temperature of the grains, causing the evaporation of volatiles and 
potentially leading to the disruption of the coagulated dust grains.  
The gas kinetic temperatures in cores toward the Orion cloud show a large drop between the ONC region and 
the L1641 region, supporting our hypothesis.

\subsubsection{Temperatures.} 
Given the low counts of our spectra and their low resolution, a detailed analysis of 
the thermal components of the emitting plasma is not possible. 
Nevertheless, a rough analysis is still possible to
differentiate between hot or cold spectra. The spectra with highest count rates 
were modeled with a two-component thermal model. In these 40 cases (cf. Table \ref{bestfit}), 
we obtained a representative mean 
temperature by calculating the weighted mean of the two temperatures, the weights being the values 
of emission measures obtained for each component in order to compare with single temperature best fit models.

Class~I objects appear on average hotter than Class II and
Class III objects. The median plasma temperatures are: $kT = 1.8\pm0.7$ keV (Class I), 
$1.2\pm0.6$ keV (Class II), $0.9\pm0.3$ keV (Class III candidates). 
The absorption due to the gas around Class I and Class II objects can mimic hotter spectra,
since the presence of any soft component below 1 keV is heavily reduced. Similar differences of
temperatures are also observed in the Rho Ophiuchi forming region \citep{Pillitteri2010aa}. 
{ In some cases, the spectra of YSOs in the Ophiuchus molecular cloud show a 6.7 keV line 
from highly ionized
Fe atoms, and this is evidence of very hot plasma ($kT\geq 5 keV$). 
In \soxs spectra we do not detect this feature, either because of the lower
count statistics or because of an intrinsically cooler temperature.    
As a consequence, we cannot disentangle the effects of a genuinely high temperature 
from those related to strong absorption.}
\subsubsection{X-ray luminosities.} 
For the brightest sources, the X-ray luminosities of classified PMS stars 
have been calculated based on fluxes
directly determined from the best fit modeling.  For the remaining YSOs, faint in X-rays, we have used the
PIMMS software to determine a count rate to flux conversion factor.
As discussed before, the values of absorption and plasma temperatures differ, on average, 
among the different SED class samples. For each sub-sample (protostars, 
Class II and Class III sources), we used a thermal APEC 1-T model, 
with $N_H$ absorption and plasma temperature corresponding to the median of values derived from the spectral
analysis of bright spectra with XSPEC program
\footnote{ Another approach that we tried was to estimate the absorption for faint sources from
the relationship between the median energy of their spectra  and $N_H$. 
\citet{Feigelson05} showed that for \chandra-ACIS COUP data this 
can be calibrated  from the spectral analysis of bright sources.
However, an attempt to use this method revealed that the relationship in \xmm-EPIC is steeper 
than in \chandra-ACIS due to the different responses. 
For median energies below 1.3 keV and a range of $N_H = 10^{19} - 3\times10^{21}$ cm$^{-2}$ 
this method is thus ineffective.}.

We compare the cumulative distributions of X-ray luminosities (CDXLs) of protostars, Class II objects and Class
III candidates in the regions around $\iota$~Ori (field $S1$), L1641 North (fields $S2+S3+S4$) and L1641 South 
(fields $S7+S8+S9+S10$). Class III candidates are selected based on their 
X-ray emission and we do not have a complete membership information for undetected Class III objects. 
Furthermore, a few foreground stars are expected in this sample. In \S \ref{xlfmem} we will calculate
X-ray luminosity function for the sub-sample of YSOs that have membership and youth information
obtained with additional optical spectroscopy and thus we can also consider those YSOs undetected in X-rays.  
\begin{table}
\caption{\label{lxmed} Median values (and mean absolute deviations, MAD, in parentheses) 
of $\log L_X$ of YSOs separated by SED class and region. { MADs are calculated as in R-language
implementation, and are scaled for the factor 1.4826 which ensures consistency with $\sigma$ of
a normal distribution. }
In the $\iota$ Ori field we detected only one Class I object.} 
\begin{center}
\begin{tabular}{l | r r r r} \hline \hline
 & Classes:	& 	& 	&	\\
 Regions       & Class I   & Class II & Class III & Class II+III \\ \hline
All     & $29.84 (0.56)$& $29.69 (0.67)$  & $29.57 (0.63)$ & $29.60 (0.64)$ \\ 
$\iota$ Ori & $ [30.23]$ & $30.06 (0.58)$ & $29.75 (0.51)$ & $29.77 (0.58)$ \\ 
L1641 N & $29.92 (0.54)$ & $29.58 (0.56)$ & $29.50 (0.63)$ & $29.51 (0.58)$ \\
L1641 S & $29.74 (0.09)$ & $29.72 (0.81)$ & $29.50 (0.69)$ & $29.55 (0.71)$ \\\hline
N1+N2+N3& $29.16 (0.70)$ & $29.93 (0.60)$ & $30.00 (0.59)$ & $29.96 (0.60)$ \\\hline

\end{tabular}
\end{center}
\end{table}

\begin{figure*}
\begin{center}
\includegraphics[width=0.89\textwidth]{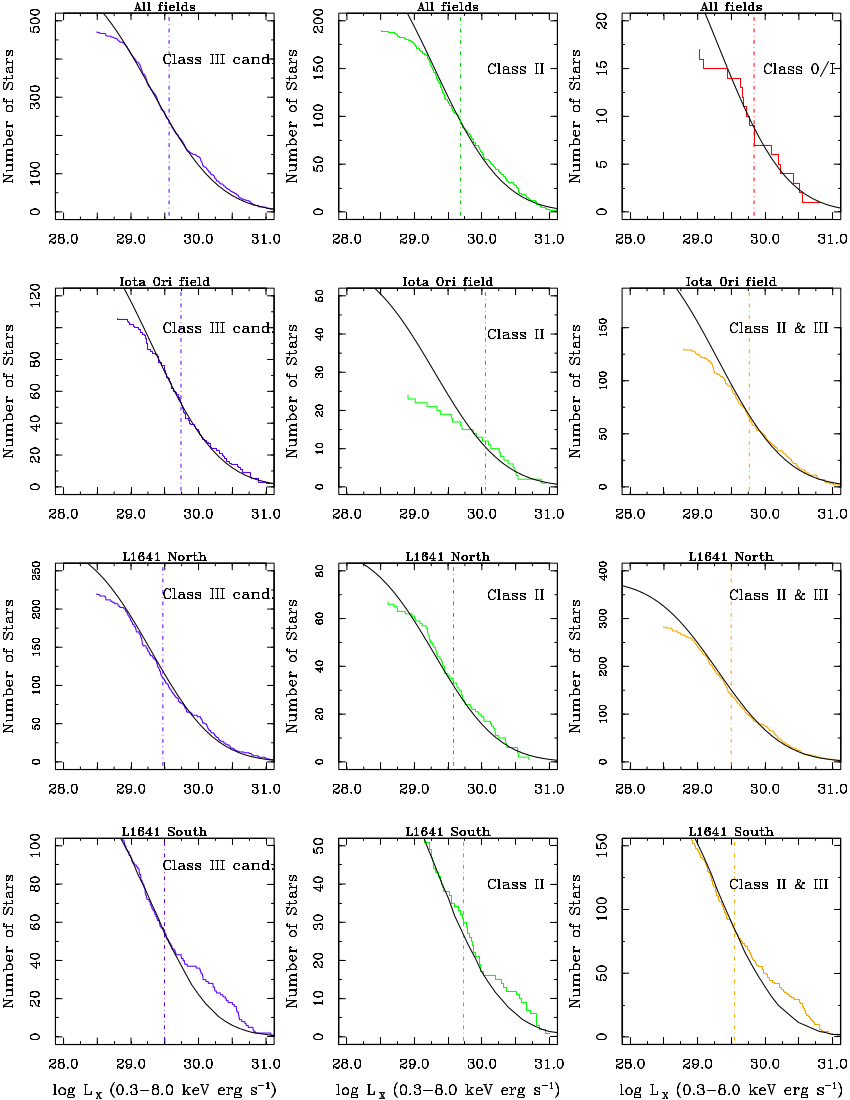}
\end{center}
\caption{\label{lx-cdf} Cumulative distributions of X-ray luminosities (CDXLs) of Class I (red), Class II YSOs (green)
and Class III candidates (blue). The top panel shows the curves for each sub-sample, 
whereas the other three panels show the curves for sub-regions: field of $\iota$ Ori, L1641~N, and L1641~S.
Vertical colored lines are the medians of the distributions. For each region we have also considered the CDXLs
of Class III candidates and Class II objects. The agreement with the model curve (black solid curve) is
good in general down to $\log L_X \sim 29.5$. The deviation from the model curve is observed in L1641N 
around  $\log L_X \sim 30$ and suggests that there is a sub-sample of stars closer than the ONC.}
\end{figure*}

In Fig. \ref{lx-cdf}, we show the CDXLs of detected YSOs separated by SED classification and by regions.  
In Table \ref{lxmed}, we report the median and the mean absolute deviation (MADs) of the luminosities 
for the YSOs separated by SED class and zones. 
The panels in the top row of Fig. \ref{lx-cdf} comprise all the detections
separated by SED Class only. 
For comparison, we plot also a log-normal model of the X-ray luminosity 
function similar to the one \citet{Feigelson05} from COUP data (solid black line).
We use this model (XLF$_\mathrm{COUP}$) to roughly fit the high luminosity tail of the CDXLs down 
to $\log L_X \sim29-29.5$. 
Below this threshold, the number of undetected sources should increase significantly, and the observed CDXL 
deviates from the log-normal model. 
{ Regarding to the adoption of the XLF$_\mathrm{COUP}$, a note of warning has to be added. 
While the existence of a universal XLF was initially proposed by \citet{Feigelson05}, subsequent 
studies have shown that the modeling and the slope of the high luminosity tail of the distribution
can depend on the masses of stars, their distribution, and their age with respect to 
the COUP sample \citep{Getman2006, Wang2008, Kuhn2010}. The modeling of the XLD of COUP has been
changed: \citet{Getman2006} used a broken power-law to model the lightly obscured 
YSOs in COUP dataset, while in other SFRs a log-normal model gave a good fit to the XLDs. 
\citet{Prisinzano2008} have calculated mass-stratified XLFs from the COUP data for different SED classes,
adopting a log-normal model.
Among Class II and III YSOs with masses in $0.1-0.9$ M$_\odot$, 
they find similar medians but different widths of XLFs, with the Class III having a
narrower distribution of luminosities (Fig. 9 in \citealp{Prisinzano2008}).}

We have used the same median of XLF$_\mathrm{COUP}$ ($\log L_X =29.3$) but we have changed the normalization 
and adjusted the standard deviation to the range $\sigma = 0.7-0.8$ for an improved fit. 
The narrower distribution suggests an intrinsic difference between L1641 and the ONC, perhaps 
due to the lack of high luminosity massive stars \citep{Hsu2012}. A similar result has been
observed by \citet{Guenther2012} in an analysis of the low mass star forming region IRAS~20050+2720.
The CDXLs of Class I, Class II and III objects are in reasonable agreement with the log-normal model for
$\log L_X \ga 29$. 

Protostars are slightly more luminous than Class II and Class III candidates 
(panels in top row of Fig. \ref{lx-cdf}). 
This over-luminosity is likely due to a selection bias because protostars are highly embedded 
X-ray sources which suffer from higher absorption and extinction when compared to Class II 
and Class III objects. The high number of undetected protostars (124 out of 147) means that 
we have detected only those that are exceptionally bright and hot in X-rays. 
The sources with fainter and/or cooler spectra are too absorbed to be detected in X-rays by our survey.

The comparison of CDLXs and the medians of $\log L_X$ of Class III candidates show 
that around {$\iota$ Ori} Class III candidates are globally more luminous  (by 0.25 dex) than  
the samples of L1641 S and L1641 N (Fig. \ref{lx-cdf}, first column plots, and Table \ref{lxmed}, third column). 
A Kolmogorov-Smirnov test gives a probability of $<0.5\%$ for the two distributions being drawn from 
the same parent distribution.
The same is not evident from the comparison of the CDLXs and the medians of Class II objects in the same
regions.

We observe an excess  around  $\log L_X \sim 30$ \lxu in the CDLX of Class II and III objects 
in L1641 N with respect to the log-normal curve.
These differences can be ascribed to the presence of two groups of stars at two different distances.
Around \iori, and to some extent in L1641 N, there is a population of  older and more evolved 
Class III stars at a closer distance than the ONC and L1641. The fraction of Class II objects
belonging to this cluster is much lower than in the ONC. 
The presence of this closer cluster causes a small excess of CDLXs of Class II of \iori\ 
and L1641 N, but produces the 0.25 dex higher luminosity in Class III sample around \iori. 
In the following section \S \ref{xlfmem} we make use of spectroscopic membership information 
to confirm this result.
\subsubsection{X-ray luminosity functions of spectroscopic members (XLF)}
\label{xlfmem} \citet{Hsu2012} published a spectroscopic survey of stars in L1641 filament.
By using H$\alpha$ emission and Li absorption at 6708\AA\ they classified 864 stars as members from a list of 
objects taken from the \spitzer\ catalog \citep{Megeath2012}. 
They suggest another 98 probable members of the region, and
estimate that the total population in L1641 is of order of 1600 stars, which is consistent with our estimate.
The area surveyed by \citet{Hsu2012} covers the region observed with \spitzer\ (cf. right panel of 
Fig. \ref{soxsmap}) 
while the \xmm\ observations cover only the more active eastern half of the \spitzer\ survey of L1641. 
Nonetheless, 652 out of the 864 members ($\sim75\%$) fall in the \soxs\ field of view.
Hsu et al. determined that 406 members have IR excess while 458 do not. 
We have considered only the conservative sample of 406+458 members of Hsu et al. (2012) in the following analysis.
We have matched our list of Class III candidates and Class II objects detected in X-rays
against the list of 864 spectroscopic members in order to determine more precisely their XLFs. 

\begin{figure*}
\begin{center}
\includegraphics[width=0.85\textwidth]{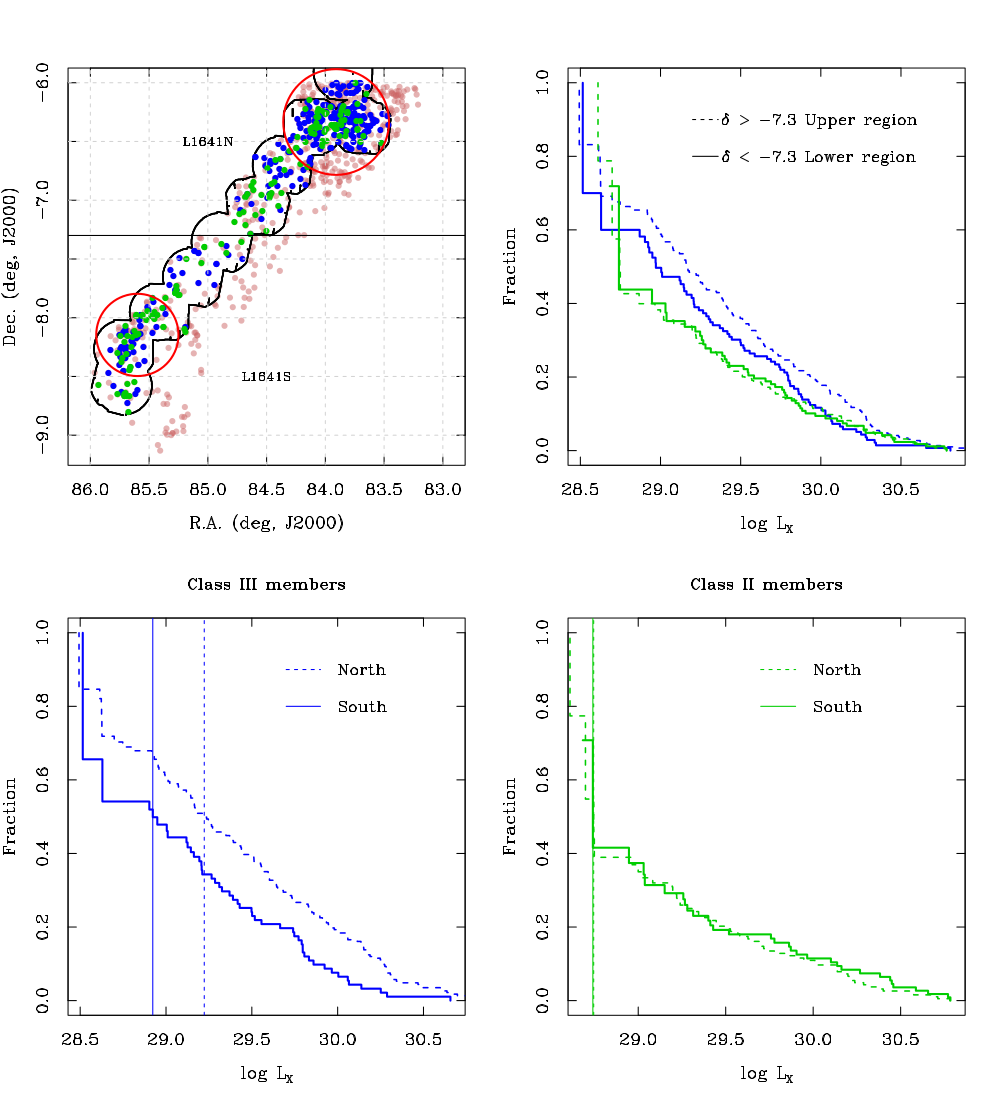}
\end{center}
\caption{\label{memxlf} Top left panel: positions of the spectroscopic members from \citet{Hsu2012} 
(red symbols), matched with Class III candidates (blue points) and Class II objects (green). 
Horizontal lines and circles mark the sub-regions used to divide the stars into L1641~N and L1641~S
and their core regions. Top right: XLFs of Class II (green) 
and Class III stars (blue) for stars north (dashed line) and south (solid line) to  $\delta = -7.3$. 
Bottom panels: XLFS for Class III (left) and Class II objects in the cores of L1641~N (dashed line) and 
L1641~S (solid line).
{ Vertical lines mark the medians of XLFs for the South (solid line) and North (dashed line) cores.}
A difference in the XLFs is markedly visible only between Class III stars belonging to L1641~S and L1641~N.}
\end{figure*}

The XLFs were computed with a version of ASURV software  \citep{Lavalley92} to take into account the
upper limits of undetected  members.
For the members not matched with Hsu et al. catalog and within the \soxs fields, 
we have calculated upper limits to the X-ray luminosity.
Figure \ref{memxlf} shows the positions of the matches between our catalog of Class II and Class III candidates
and the \citet{Hsu2012} list of members (top left panel). The same figure shows the XLFs of the stars divided 
roughly into two groups, above and below $\delta=-7.3$ (top right panel), 
as well as those in the cores of L1641 North and South respectively (bottom panels).
We have divided the stars in two spatial groups to test systematic differences between the northern and the southern
parts. The Class III stars exhibit a large difference in X-ray luminosity when comparing those above and 
below  $\delta=-7.3$, and the difference is even more marked when considering only the core regions with
the highest stellar density in L1641 
North and South. Only Class III stars exhibit such strong difference in the respective XLFs. 
Class II objects have a similar XLF in both spatial subgroups.
The difference of medians of the XLFs of the cores between L1641~N and L1641~S is $\sim0.3\pm 0.1$ dex. 
{ When comparing with the XLFs of Prisinzano et al. (2008) in the same range of masses, 
we observe that the range of the masses, the medians for the XLFs of the Class III stars around \iori\ 
are lower than those reported by Prisinzano et al. of $\sim0.3$ dex.}
{ The  stellar masses of the members of Hsu et al. catalog with X-ray detection are in the range $0.1-0.6$,
which is one of the mass-strata defined by \citet{Prisinzano2008} to calculated their XLFs from the COUP dataset.
The difference among Class III stars is best explained by arguing that there is a group of Class III 
stars in L1641~N closer than the ONC.} 
The fact that Class II objects do not show the same difference implies also that the closer cluster 
is composed mainly of stars older and more evolved than those in the ONC but still in the PMS phase. 
The same conclusions have been drawn by \citet{Alves2012}, using infrared photometry and optical spectroscopy 
to determine the membership and the characteristics of YSOs in Orion~A.

\subsection{Comparison with the ONC \label{onccomp}} 
We have analyzed the observations available in the \xmm\ archive and listed them in Table \ref{fields} 
in order to compare the X-ray emission in L1641 with that of the young stars around the densest part of the ONC. 
The core of the ONC is located between fields $N2$ and $N3$. 
Fig \ref{map_north} shows the positions of protostars, Class II objects and Class III candidates in the
northern fields. A strong alignment of protostars in a filament running along of the three
northern fields is evident. Also, the Class II objects are especially concentrated around the core of the ONC, 
while Class III candidates are more spread out. 
The ratio between the number of Class III candidates and Class II objects detected in X-rays 
decreases from north to south. This ratio is $\sim 2.2, 0.9$ and 0.52 in $N1$, $N2$, 
and $N3$ fields, respectively. 
The increasing number of Class II objects with respect to the number of disk-less stars can be understood 
as an evolutionary effect because younger and less evolved stars are found closer to the core of the ONC, 
while more evolved stars are spread at larger distances. 

\begin{figure}
\begin{center}
\includegraphics[width=0.99\columnwidth]{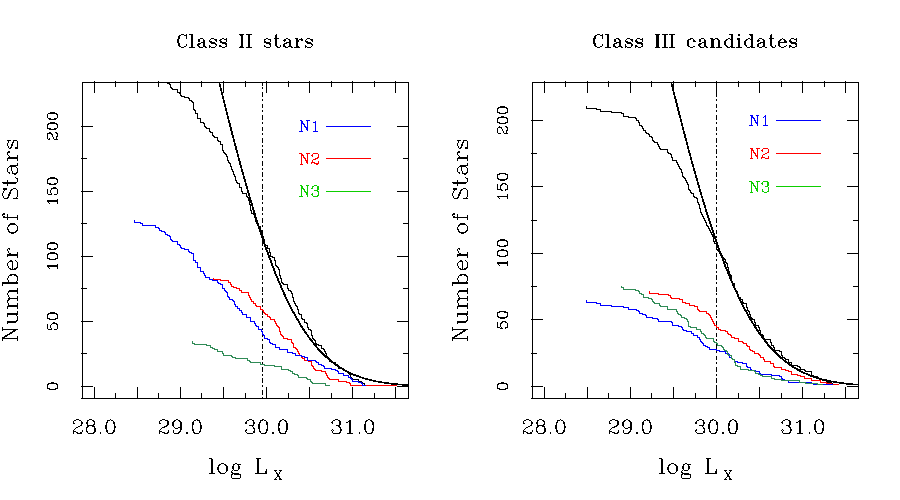}
\end{center}
\caption{\label{xlf_north}XLFs of Class II objects and Class III candidates in the northern fields 
as in Fig. \ref{lx-cdf}. The solid stepped curve is the total sample, colored curves are the samples 
in the three northern fields as indicated by the labels. 
The dot-dashed vertical lines mark the medians of the total sample of Class II and Class III
stars. The smooth curve is the model distribution.}
\end{figure}

\begin{figure}
\begin{center}
\includegraphics[width=0.49\columnwidth]{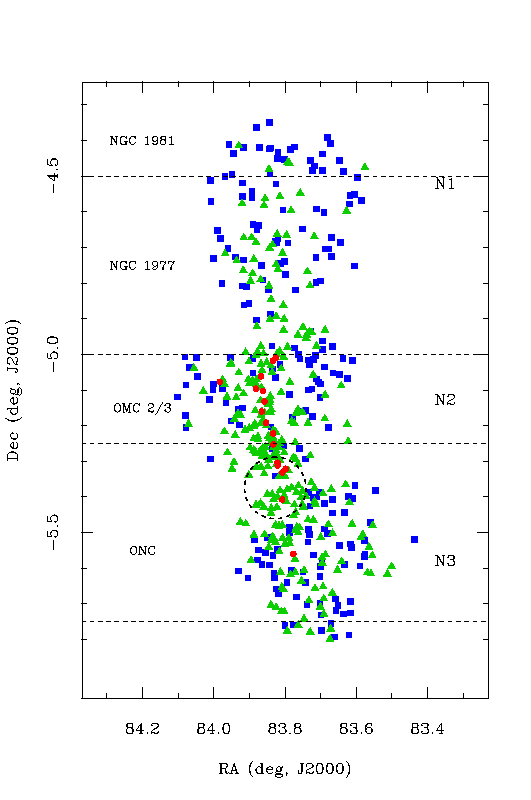}
\end{center}
\caption{\label{map_north} Map of the YSOs detected in X-ray in the northern fields $N1$, $N2$, $N3$ 
around ONC. We marked with colored large symbol the sources matched with \spitzer\ YSOs \citep{Megeath2012}.
Red circles are protostars, green triangles are Class~II stars, blue squares are Class~III candidates.
The main  sub-regions are separated by horizontal lines. The crowded region excluded from X-ray data is shown
with a dashed circle. Class~II objects and protostars preferentially lie along the gas filament, 
while the Class III candidates are more spread out.}
\end{figure}

We have derived X-ray luminosities distribution (CDLXs) for X-ray detected Class II and Class III candidates
and compared these with  L1641 YSOs. 
We have excluded the sources located near the core of the ONC in a circle
of center RA: $05^h35^m19.3^s$, Dec: $-05^d21^m47.6^s$ and radius 5.2\arcmin\ that corresponds to
0.6 pc at the distance of the ONC, and encompassing its densest part (marked in Fig. \ref{xlf_north}). 
This is a region with many confused and blended sources in the \xmm\ image (Fig. \ref{northernfields}),
and the instrument has a less calibrated {\em PSF} because it is at the edge of the field of view.
Fig. \ref{xlf_north} shows the XLFs of Class II objects and Class III candidates detected in the 
northern fields N1, N2 and N3.

From the fits of CDLXs to the log-normal model,  we infer a normalization 
of 550 stars for the sample of Class II objects and Class III candidates. This, combined
with the number of detections in each sample (235 and 211 objects), suggests a rate of
detection of $38-42\%$ and thus a total population of $\sim 1200\pm 100$ YSOs in these
fields (excluding the core of the ONC). 

The number of detected Class II stars (235) is slightly larger than that of Class III objects (211).
The flatter slope of Class III CDLXs at low luminosities suggests that a fraction of Class III objects 
could be undetected in this range of luminosities. 
The ratio of Class III to Class II objects could be close to 1, lower 
than around \iori (cf. Table \ref{det_table}). Interpreting this ratio as an { evolutionary tracer},
it suggests that this group of stars is younger than the stars around \iori.  

\subsection{\label{spacedensity} Spatial distribution of PMS stars}
\begin{figure*}
\includegraphics[width=\textwidth]{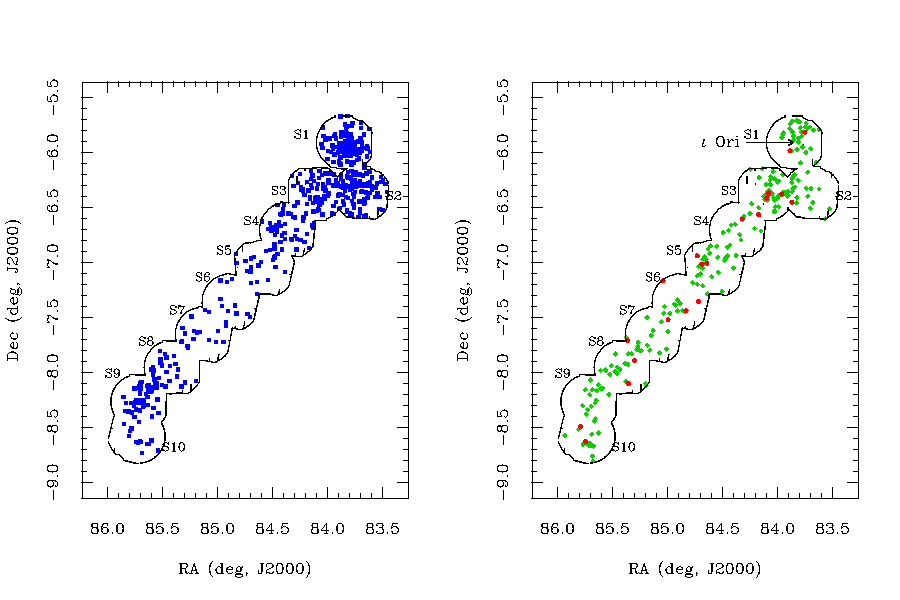}
\caption{\label{map-yso}
Distribution of Class III candidates (left panel) and Class II YSOs detected in X-rays (right panel, 
green symbols) toward L1641. Red filled circles in the right panel are the protostars detected in X-rays.
Class III candidates are grouped in two main sub-clusters, one around \iori\ and L1641~S. 
Class I and Class II objects are detected along the gas filament and less spread out than 
Class III candidates.}
\end{figure*}

In this section, we give a description of the spatial distribution of the 
young stellar population around \iori\ and in L1641, based on the classifications of YSOs
obtained with \spitzer\ and \xmm\ data. 
The analysis in \S \ref{abs} shows that, on average, Class III objects are found in less absorbed regions 
(see Fig. \ref{nh-av}) and that they could be a slightly intrinsically more X-ray 
luminous than Class II objects (see also \citealp{Preibisch05}). 
This can lead to an easier detection of Class III stars with respect to Class II objects in X-rays. 
Fig. \ref{map-yso} shows the positions of the Class III candidates (left panel) and Class II 
and protostars detected in X-rays (right panel). 
In both samples a dominant clustering of YSOs appears around  \iori\ and near L1641~S,
while the central region of L1641 is relatively less populated.

\begin{figure*}
\begin{center}
\includegraphics[width=0.79\columnwidth, angle=0]{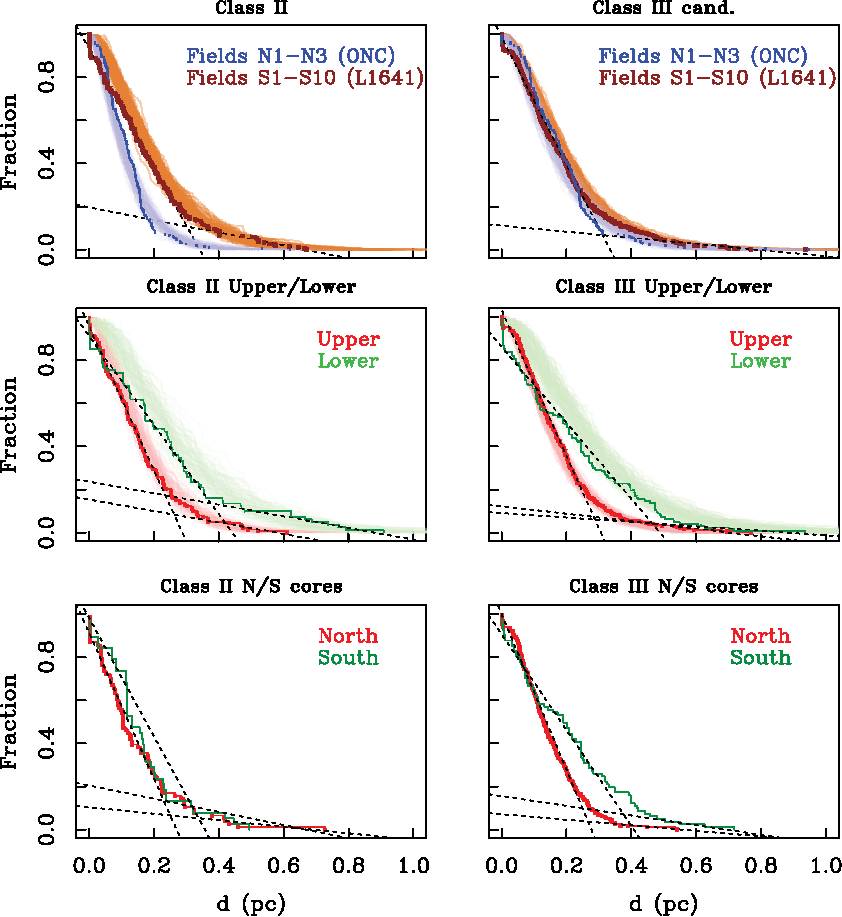}
\end{center}
\caption{\label{nnlengths} 
Fractional cumulative distributions of nearest neighbor distances of Class II objects 
and Class III candidates. Top panel: distributions for the samples of Class II objects (left panel) 
and Class III candidates (right panel) in L1641 and ONC.
Central and bottom panels are the same for the samples in the upper and lower regions of L1641 and
their cores. Dashed lines are the linear fits to the cores and the tails of the distributions. 
Shaded areas show an area encompassing the 10,000 Monte Carlo simulations of uniform distributions.}
\end{figure*}

To address the question of how clustered are the YSOs in our survey, 
we use the distributions of distances to the nearest neighbor 
(NN distances) and Monte Carlo simulations. 
For each object in each sample of Class II and Class III YSOs, we calculated the distance to the closest
object and we obtained a cumulative distribution of distance greater than a certain value, 
and normalized for the total number of objects in that sample. We obtained analogous distributions
for the ONC, the upper and lower region of L1641, their cores as defined in \S \ref{xlfmem}, and the ONC. 
These distributions are shown in Fig. \ref{nnlengths}. 
The number of objects in each sample can bias the comparison of the NN distributions. 
We have scaled the NN distances of the Class II samples to take into account the different numbers of objects
in the compared samples.

The comparison of the observed distribution and the 
envelope of the simulated distributions of NN distances will show any differences with respect 
to a uniform distributions of stars, and quantify any local clustering of objects 
within a spatial scale significantly smaller than the surveyed area. 
For this purpose, we ran a set of Monte Carlo simulations of uniformly distributed stars 
with size equal to the total sample of L1641 Class II and Class III objects and that of the ONC samples. 
The curves of observed NN distances are not fit by a gaussian model. 
Rather, a linear fit to the cores of the distributions and their wings, similar to the 
approach followed by \citet{Gutermuth09}, allows us to estimate a characteristic separation
within which most of the objects are grouped.

In the ONC, the distributions of NN distances show a concentration of objects at $d<0.2$ pc 
while L1641 shows the bulk of NN distances within 0.3 pc (top panels in Fig. \ref{nnlengths}).
NN distances of Class II and III objects in L1641 are marginally consistent with an 
uniform distribution. These are  more discrepant at short distances, 
while the distribution of objects further out more closely follows a uniform distribution. 
At NN distances shorter than 0.15 pc the agreement with the uniform distribution is at level  of $\le1\%$,
thus suggesting that the stars in both samples of Class II and III objects are spaced more closely than expected from a uniform distribution. 
This is even more evident in the sub-samples of the upper/lower regions of L1641 and in their respective cores
(See Fig. \ref{memxlf} for definition of cores).
Class III stars in the lower region and the southern core are markedly different from an uniform 
distribution, while their distribution in the northern half of L1641 and the northern core 
is indistinguishable from a uniform distribution.
The distributions in the lower region as well as the southern core appear to be bi-modal, 
with a minor fraction of clustered sources with separation $<0.15$ pc, surrounded by a more dispersed 
population. The northern region shows a different situation, with the stars densely packed
and their distribution is not too different from an uniform population.
This can be seen as a bias of the X-ray survey, because the actual spatial scale of this group 
of Class III stars around \iori\ and L1641N is of the same scale or larger than the \xmm\ field of view, 
hence the stars are almost uniformly distributed within the field of view.

The presence of clustering in the two core fields can also be inferred by the typical separations of 
the Class II objects and Class III candidates in this regions.  In the cores, about 90\% of the Class 
III and Class II objects have separations < 0.3 pc. In comparison, for the entire L1641 region, 
90\% of of the Class III candidates and Class II objects have separations < 0.4 pc. 
The smaller separations in the two core regions suggest that the young stars are more clustered in these regions.
In the ONC, we find evidence of clustering by comparing the separations of the Class II and Class III objects. 
Although we excluded the densest part across $N2$ and $N3$ fields, we find a median separation of ∼ 0.2 pc 
for the Class II objects only. In contrast, the Class III objects in the ONC are more dispersed and have 
NN distances similar to those in L1641, i.e. median separations of 0.3 pc.

\begin{figure*}
\includegraphics[height=1.00\textwidth, angle=-90]{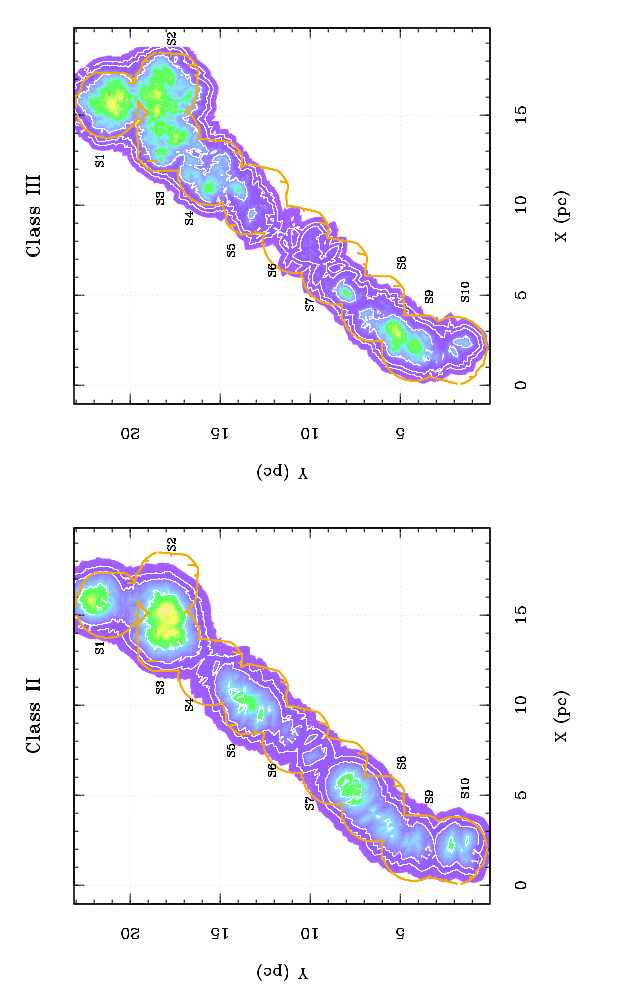}
\caption{\label{surfdens}
Maps of surface density of Class II YSOs (left panel) and Class III candidates detected in X-rays (right panel). 
White contours correspond to 25\%, 50\%, and 90\% quantiles. { The maximum surface density is about 20 pc$^{-2}$
for Class II YSOs and about 80 pc$^{-2}$ for Class III candidates.} 
The field of view of \soxs\ is marked with { an orange contour}. 
The density maps show { an imprecise match} between the peaks of surface densities of the Class II objects 
and the Class III stars, suggesting that star formation toward the Orion~A cloud is not coeval,
the presence of multiple events of star formation along the line of sight of Orion~A, and/or migration
of the older stars.}
\end{figure*}

While the distributions of NN distances enable us to determine the degree of clustering, 
the maps of surface density of stars allow us to evaluate difference in the spatial distributions
between Class II objects and Class III stars.
Fig. \ref{surfdens} shows surface density maps of Class III candidates and Class II objects.
For each point in a grid with resolution $\sim20\arcsec$, we calculated the distance to the sixth nearest
object in the sample of Class II and Class III YSOs. 
The area of the circle with the radius equal to the sixth object and the number of 
objects  inside that area are used to compute the surface density.
We have normalized the maps to the ratio of number of Class III stars to Class II objects.
In dealing with  3-D projection effects, 
these maps and the NN distances analysis are constructed by assuming that
the differences along the line of sight are similar to those in the plane of the sky.
Effects of 3-D projection could be important, and a significant fraction of the stars around 
Iota Ori can be ~100 pc closer than the ONC, as we deduced by XLFs comparison. 
These facts can constitute a bias to the NN distances analysis and to the surface density maps. However,
we cannot select the stars belonging to the foreground cluster only, and separate from the 
genuine Orion A population, although we know that a large fraction of the Iota Ori cluster 
is made by Class III stars. 
{ Being this cluster at a distance closer than the ONC, it appears denser than the
background stellar population of Orion~A.}

The Class III objects are denser in the northern part of the survey around $\iota$ Ori
and sparse in the filament of L1641. In L1641~S, Class III objects are denser across
the fields $S8-S9$. 
The density of Class III objects reaches a peak of 80 stars pc$^{-2}$ in field $S1$ (\iori), 
other local maxima are around 10$-$15  stars pc$^{-2}$. 
Hence we can recognize two main groups of Class III objects where their surface density is
above 10 pc$^{-2}$; one in the northern fields near \iori\ and the other in L1641S separated by
a very low density region centered on field {\em S6} (Fig. \ref{surfdens}, right panel). 
The absence of Class III candidates in the middle part of 
the L1641 filament is not related to a lower sensitivity of the survey in that region. 
$S6$ field has the same exposure time of other fields, here there is a real deficiency of YSOs. 

Class II YSOs are most numerous around $\iota$ Ori and merge into the ONC to the north. 
They are grouped into small sub-clusters along the filament of L1641. 
The surface density of Class II objects has a peak of $\sim$20 stars pc$^{-2}$ in L1641~N (fields $S2-S3$) 
and several  other local maxima of $4-8$ stars pc$^{-2}$ across fields $S7-S9$. 
Class II objects are distributed differently than the Class III candidates.  
By comparing the two maps (Fig. \ref{surfdens}), we notice that the high density zones do not overlap exactly. 
This is most evident in the fields {\em S2} and {\em S9}. Class II are more concentrated and offset with 
respect to the centers of the corresponding Class III clusters, suggesting a segregation effect of the YSOs 
due to their evolutionary stage.

\begin{figure*}
\begin{center}
\includegraphics[width=0.95\columnwidth, angle=0]{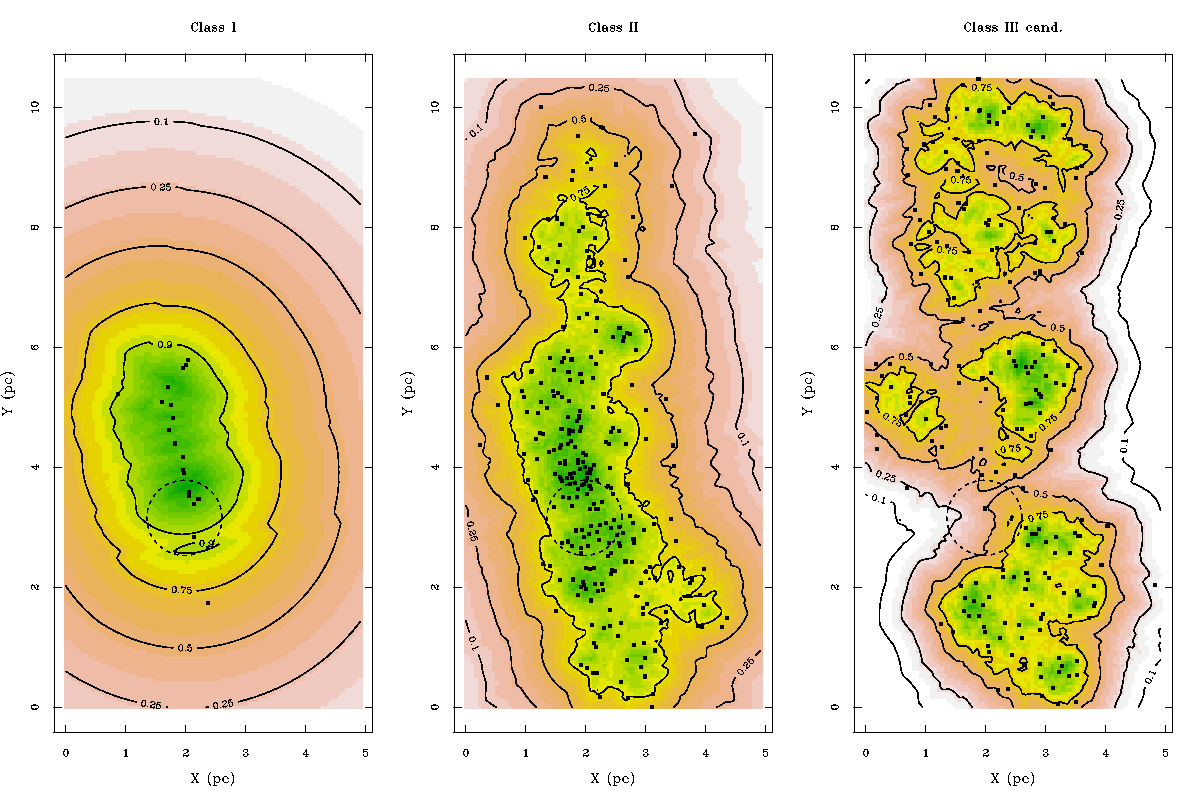}
\end{center}
\caption{\label{densnorth}
Maps of surface density of X-ray detected Class I objects (left panel), Class II objects
(central panel) YSOs and  Class III candidates (right panel) toward the ONC. 
The coordinates are given in parsecs calculated at the distance of the ONC. 
{ The maximum surface density is about 175 pc$^{-2}$
for Class II YSOs and about 60 pc$^{-2}$ for Class III candidates.}
Contour levels are traced at 10\%, 25\%, 50\% and 75\% 
quantiles (for the Class I panel also the 90\% quantile is plotted). The dashed circle marks the part of ONC
excised from XLFs analysis (cf. \S \ref{onccomp}). The filamentary distribution of Class II objects and
protostars is evident, while Class III stars are more spread out. Star formation might have proceeded 
sequentially from outside inward, or the older stars have migrated farther away.}
\end{figure*}

We have calculated the surface density maps for the YSOs in the northern fields $N1$, $N2$ and $N3$ 
(Fig. \ref{densnorth}).
These density maps are similar to the distribution of stars found in L1641 and \iori. In particular, we
notice that the Class III candidates are spread across the image but show a lower density in fields $N1$ and $N2$.
The latter includes the densest part of the ONC, where strong absorption that 
limits the X-ray sensitivity and the bright nebulosity that limits the IRAC sensitivity.
The low density in $N2$ is likely due to a real feature in the distribution of Class III stars. 

The filamentary shape of the Orion~A cloud, as traced by the positions of Class II objects and the 
few Class I protostars detected in X-rays, is evident. 
Class I objects detected in X-rays form a chain, most of them lying in the densest part of the cloud. 
The surface density of Class II YSOs at the 75\% contour is about 2 pc
wide which is comparable to the Jeans length of a cloud at T $\sim 10-15$K and mass 
of a few dozens of solar masses.   
It is worth noting the complementary nature of the distributions of Class II and Class III objects 
which is also observed  in L1641.
They show a segregation effect: Class III stars tend to occupy peripheral 
regions of the cloud. Either they have migrated from their birth place at the center of the filament, 
or, because of the collapse of the cloud, they formed first in the outer parts and then continued 
to form stars  in the central part of the filament.

\section{Identification of three Class III rich clusters towards the Orion A cloud.}
\label{newclusters}
One of the primary results of the SOXS survey is the discovery of three regions in the Orion~A 
cloud rich in Class III objects.   
The first of these clusters is toward NGC~1981 in the $N1$ field (Fig. \ref{map_north}). 
This field is distinguished by a Class III/Class II ratio of 2.2, as compared to ratios less than 
1 for the $N2$ and $N3$ fields (Sec 4.2).   
NGC~1981 is found directly north of the Orion A cloud and the NGC~1977 nebula suggesting that it formed from 
the Orion~A cloud  and has dispersed the cloud material surrounding the cluster 
(see Fig. \ref{northernfields} in this paper, Figs. 1 of \citealp{Alves2012}).  
The NGC~1977 cluster appears younger than NGC~1981 since it is richer in Class II objects  than the latter 
(Fig 10,  \citealp{Peterson2008}).   

The second new cluster of Class III stars is L1641-S located in the S9 and S8 fields toward the southern end of L1641 
\citep{Strom1993} (see Fig. \ref{map-yso}).  
This cluster has  $N_H$ values corresponding to $A_V$ values up to 18~mag (Fig. \ref{nh-av}) 
and shows an elongation aligned with the L1641 cloud, indicating that this cluster is associated with 
the  cloud.  
An HR diagram of this cluster shows an age of $\sim 3$~Myr, older than that of the other groups in the L1641 cloud 
\citep{Allen1995,Allen2008}.  The  Class III/Class II ratio of the $S9$ field where the cluster is centered, 
$r \sim5.5$, is significantly higher of that of the neighboring $S8$ and $S10$ fields, 1.0 and 0.87, respectively; 
this is a further evidence that the L1641-S cluster is significantly older than young stars in the nearby regions 
of L1641.

The  third cluster is found toward the O9III+B1III system of \iori\ and the NGC~1980 cluster 
(hereafter we refer to this as the $\iota$ Ori cluster). 
This cluster is identified by \citet{Alves2012} (see also \citealp{PillitteriAAS2011} and \citealp{Bally08}),
who noticed a concentration of low extinction stars surrounding $\iota$ Ori. On the basis of the low
extinction to the stars, they argued that this is a large cluster in the foreground of the Orion A cloud.  
Our XMM survey confirms their result by revealing an extended cluster of 
Class III objects with a density peak centered on $\iota$ Ori.  The cluster appears to extend 
through $S1$, $S2$, $S3$, $S4$  and $S5$ fields, more than 1$^o$ on the sky.   
The morphology and full extent of this cluster cannot be determined from our survey given that it appears to continue 
beyond our \xmm\ fields. On the other hand, Class II objects follow the filamentary structure of 
the molecular gas and appear to be mostly part of the molecular cloud. 

Class III objects toward $\iota$ Ori have low $N_H$ 
absorption (Fig. \ref{nh-av}), consistent with the location of this cluster in foreground respect to the Orion~A cloud.  
As shown in Sect. \ref{xlfmem}, the XLF of the Class III objects towards the $\iota$ Orion cluster is shifted by 
$\sim0.3\pm0.1$ dex to higher luminosities than that of Class III objects in the southern L1641 fields.  
This shift is consistent with the presence of a large cluster of Class III objects in the foreground of Orion~A.  
Furthermore, the shift of 0$.3\pm0.1$~dex implies a distance that is $\sim1.4$ times closer than the 
objects in the L1641 fields. If we use the adopted distance of 414 pc for L1641 
\citep{Menten07} and the ONC, then the distance to the \iori\ cluster is $\sim300$ pc and comprised 
in the range $\sim261-328$ pc.  
In contrast, the XLFs of the Class II objects coincident with the \iori\ cluster and the those in the southern region 
of L1641 are very similar, indicating that the Class II objects throughout the survey are at a common distance 
and  are associated with the Orion~A cloud. We can give an estimate of the number of members in this cluster. 
We have 313 Class III stars in fields $S1$ through $S4$. If we assume a circular shape and a 
radius of $\sim1\deg$ ($\sim5.5$ pc at 300~pc) centered on \iori, 
we estimate that about 1000 members are found this cluster. 

\citet{Hoogerwerf2001} traced the origin of the runaway O stars $\mu$ Col and AE Aur to an ejection event originating 
at $\iota$ Ori system 2.5~Myr ago.  Using Monte-Carlo simulations based on the  {\em Hipparcos} proper motions and parallaxes 
of these three stars, they determined a distribution of likely distances for the event. Their distribution peaked at 325 pc, 
although it is extended from 250 to 500 pc.  Since in 2001 the ONC was the only known nearby cluster, they proposed that 
the ONC at $\sim$414 pc was the progenitor cluster of $\iota$~Ori and the two runaway stars.  
On the basis of this hypothetical connection between $\iota$ Ori and the ONC, \citet{Tan2006} 
argued that the ONC must be a minimum of 2.5~Myr old to have had time to form the four stars participating 
this ejection event. However, the detection of this previously unknown cluster around $\iota$~Ori reveals 
the progenitor of the \iori\ system.
We note that the crude distance estimate we obtained from the XLF, $\sim300$ pc, is close to the peak of the 
distribution of  distances determined by \citet{Hoogerwerf2001} for \iori.  
Furthermore, given that the cluster is primarily traced by Class III objects, we argue that the $\iota$ Ori cluster 
is $\ge 5 Myr$ old, consistent with the 4-5 Myr age estimated by \citet{Alves2012}. This is also consistent with the age 
of the ejection event of $\mu$ Col and AE Aur.   The association of the ejection event with the $\iota$ Ori cluster 
also removes the minimum age constraint of the ONC ($t \ge 2.5$ Myr) given by \citet{Tan2006}.  

\section{Conclusions} \label{conclusions} 
In this paper, we have presented the initial results of the SOXS survey, a {\em Survey of Orion A with XMM-Newton 
and Spitzer}, designed to explore the young stellar population in the Orion A molecular cloud south of the Orion Nebula, 
typically referred to as the Lynds 1641 region (L1641). 
L1641 is a filamentary and highly structured part of the Orion~A cloud that extends southeast of the ONC and fills 
a  $3^{\circ} \times 1^{\circ}$ region. The X-ray survey focused on the eastern denser half of the cloud. 
We have used a multi-band approach for studying the pre-main sequence stars and protostars, 
which were identified either by IR excesses in the Class I and II phases, or by elevated X-ray emission 
through to the Class III stage. We have described the data analysis and the main properties that we have derived 
from X-rays observations. We have analyzed the spectra of the brightest sources, derived fluxes and luminosities, 
obtained X-ray luminosity functions (XLFs) of the PMS stars in the survey. 
Furthermore, we have derived spatial density maps of the young stars and compared their spatial distribution in 
L1641 to that of the ONC.

We have detected 1060 X-ray sources, with 972 sources having at least one Spitzer counterpart, and about 70\% of 
these being young stars in evolutionary stages ranging from protostars to disk-less young stars. 
We estimate that are between 1800 and 2350 young stars in SOXS.  We further estimate that there may be $\sim365$ 
more young stars when extrapolating to the eastern L1641 fraction surveyed by Spitzer alone. 
This population of young stars is comparable or larger than the population in the ONC, yet 
it is dispersed in a larger area and contains very few massive stars.

Through simple 1-T and 2-T APEC absorbed models, we have estimated absorption, plasma temperatures and fluxes for 
232 young stars and protostars.  We find that the plasma temperatures depend on Class, with Class I objects 
(i.e. protostars) being hotter than Class II objects (pre-main sequence stars with disks) 
and Class III candidates (disk-less pre-main sequence stars). $N_H$ values derived from the 
X-ray spectra map the gas absorption to the young stars. A gradient of $N_H$ from north to south is observed, 
with a large number of less absorbed Class III candidates around \iori\ and in L1641 North, whereas the heavily 
absorbed Class I/II YSOs are located primarily near the southern and the central parts of L1641. The $N_H /A_K$ 
for L1641 ratio is found to be lower than the average value found in the ISM or high mass star forming regions, 
but consistent with what is found in NGC 1333 and Serpens. While high mass star forming regions, such as RCW~108 and 
RCW~38, show ISM-like $N_H /A_K$ ratio, here we find less gas absorption with respect to dust extinction than 
typically seen in the ISM. An explanation for this behavior is that low mass star forming regions lack intense 
radiation fields typical of high mass star forming regions. We speculate that the ambient radiation field may 
influence the coagulation and the sizes of dust grain and thereby the $N_H /A_K$ ratios.

We have analyzed the spatial distribution of the pre-main sequence stars and protostars. Class I and II objects 
appear clustered into small groups in the L1641 North and South regions. The Class III candidates are in 
contrast more dispersed with two main groups around \iori\ and in L1641~S. In the central fields of L1641, 
the Class III stars are less common. We suggest that the Class III objects preferentially trace an older population 
of pre-main sequence stars towards L1641. Strong evidence for this older population is found in a cluster of 
Class III stars around the massive B0III star O\iori\ (the southernmost star of Orion’s sword) and NGC 1980. 
The high concentration of Class III stars in this region further confirms that the cluster of low extinction stars 
found by Alves \& Bouy (2012) around \iori\  cloud is a cluster of pre-main sequence stars that formed prior to the ONC. 
We also find a concentration of Class III objects toward L1641~S, this is the first evidence for a cluster of older 
stars in the L641 dark cloud itself.

We have compared the X-ray luminosity functions (XLFs) of Class II objects and Class III stars for which the membership 
has been determined on the basis of combined criteria of IR excess (for Class II objects), X-ray emission (for Class III 
stars) and/or the presence of spectral features such as Li doublet absorption or H$\alpha$ emission (Hsu et al. 2012). 
The XLFs of Class III stars reveal that the group of stars in L1641N and around \iori\ appears more luminous than those in L1641S. We ascribe this to a closer distance ∼ 300 pc for the cluster of stars surrounding $\iota$ Ori.  
These findings indicate that the cluster of stars of which \iori\ is the most massive remaining member, 
formed in the foreground of the ONC. Furthermore, the cluster surrounding $\iota$ Ori appears to be the site of 
the proposed ejection of the runaway O stars $\mu$ Col and AE Aur $\ge2.5$ Myr ago.  

We have in addition analyzed three archival XMM fields covering the northern Orion~A cloud. 
The X-ray sources in this northern regions follow the narrow filamentary structure of the ONC. 
Class II and Class I objects are detected along the cloud filament, while Class III candidates are spread out 
along the peripheral borders of the filament. We suggest that the Class III candidate are more spread out because 
they have had time to drift away, or that the they formed by a sequence of star formation progressing inward toward 
the filament.  We also find a Class III rich cluster of young stars towards NGC 1981; this appears to be 
the result of a previous star formation event directly north of the Orion~A cloud. 
In total, the SOXS survey has revealed the non-coeval nature of the star formation toward the Orion A.  
Not only do we find further evidence for a foreground cluster surrounding $\iota$ Ori, but we also find more evolved 
regions of recent star formation at both the northern and southern ends of the Orion A cloud.  
This non-coevality is an important constraint on models for the formation and evolution of the Orion~A cloud.  

\acknowledgments

The {\em XMM-Newton} guest investigator program supported IP through grant NNX09AP46G.
S.J.W. was supported by NASA contract NAS8-03060 to the Chandra Science Center.
Based on observations obtained with XMM-Newton, an ESA science mission with instruments and contributions 
directly funded by ESA Member States and NASA.
This work is based in part on observations made with the \spitzer\ Space Telescope, 
which is operated by the Jet Propulsion Laboratory, California Institute of Technology under a contract with NASA.
This publication makes use of data products from the Two Micron All Sky Survey, 
which is a joint project of the University of Massachusetts
and the Infrared Processing and Analysis Center/California Institute of
Technology, funded by the National Aeronautics and Space Administration
and the National Science Foundation.
{\it Facilities:} \facility{\xmm\ (EPIC)}, \facility{\spitzer\ (IRAC+MIPS)}.

\clearpage

\end{document}